\documentclass{article}
\usepackage{fancyhdr}
\usepackage{amscd}
\usepackage{graphicx}
\usepackage{amsmath}
\usepackage{amssymb}
\usepackage{amsthm}
\usepackage{mathrsfs}
\newtheorem{definition}{Definition}[section]

\newtheorem{remark}[definition]{Remark}

\numberwithin{equation}{section}

\newcommand{\g}{\text{\boldmath $g$}}
\newcommand{\I}{\text{\boldmath $I$}}

\newcommand{\n}{\text{\boldmath $n$}}

\renewcommand{\u}{\text{\boldmath $u$}}

\def\t{\text{\boldmath $t$}}

\def\R{\mathbb R}

\begin{document}

\title{A mathematical treatment of bump structure for the particle laden flows with particle features}

\author{Kaname Matsue\thanks{Institute of Mathematics for Industry, Kyushu University, Fukuoka 819-0395, Japan {\tt kmatsue@imi.kyushu-u.ac.jp}} $^{,}$ \footnote{International Institute for Carbon-Neutral Energy Research (WPI-I$^2$CNER), Kyushu University, Fukuoka 819-0395, Japan}, Kyoko Tomoeda\thanks{Institute for Fundamental Sciences, Faculty of Science and Engineering, Setsunan university, Neyagawa Osaka, 572-8508, Japan {\tt tomoeda@mpg.setsunan.ac.jp}}
}
\maketitle

\begin{abstract}
In this paper, we consider the particle laden flows on a inclined plane 
under the effect of the gravity. It is observed from preceding experimental works that the particle-rich ridge is generated near the contact line. 
The bump structure observed in particle-rich ridge is studied in terms of Lax's shock waves in the mathematical theory of conservation laws.  
In the present study, the effect of particles with nontrivial radii on morphology of particle 
laden flows is explicitly considered, and dependence of radius and concentration of particles on the bump structure is extracted. 
\end{abstract}


\section{Introduction}
\label{intro}
\par 
In this paper we focus on particle laden flows on a inclined plane under the effect of the gravity. 
Particle laden flows stem from fluids including plenty of particles playing dominant and considerable roles in dynamics, 
such as debris flow, slurry transport, and mixing of pharmaceuticals {\cite{ZDBH}}.  
Zhou et al. {\cite{ZDBH}} have presented experimental results for the behavior of particle laden inclined film flows.  
Their experiment consists of an acrylic plate with an adjustable inclination angle $\alpha$ ($0^{\circ}<\alpha <90^{\circ}$).  
A particle laden of polydisperse glass beads (diameter 250-425 mm) flows down on an acrylic plate, Zhou et al. {\cite{ZDBH}} 
have observed three different particle behavior depending on the inclination angle and particle concentration as follows: 
(a) At low inclination angles and the particle concentration, the particles settle to the substrate and the clear silicone oil flows over the particle bed. 
(b) At high inclination angles and the particle concentration, the particles accumulate at near the contact line forming a particle-rich ridge. 
(c) At intermediate inclination angles and the particle concentration, the particles remain well mixed in the fluid. 
\par 
Zhou et al. {\cite{ZDBH}} have presented a lubrication model to capture the behavior of the particle-rich ridge regime 
as observed in the ridged regime (b). 
To derive the lubrication model, they have considered the governing equations of motion for the mixture. 
The governing equations are given by the following system of partial differential equations for the particle concentration 
$\phi(x, t)$ ($0\leq \phi(x, t)  \leq \phi_m$) and the velocity of particle laden $\u=(u, v)$; 
\begin{eqnarray}
-\nabla\cdot\left(-p\I+\mu(\phi)(\nabla\u + \nabla\u^{T})\right) &=&\rho(\phi)\g, \label{NS1} \\
\nabla\cdot\u&=&0, \hspace{0.5cm}\hbox{ in }\hspace{0.3cm}0 < y < h(x, t), 
\hspace{0.3cm}t\geq 0, \nonumber  
\end{eqnarray} 
where $h(x, t)$ is the total thickness of particle laden, $\g= g(\sin \alpha, -\cos \alpha)$ is the acceleration of gravity, $p$ is pressure and $\I$ is the $2$-dimensional identity matrix.  
$\mu(\phi)$ is the effective viscosity given by $\mu(\phi)=\phi_m^2/(\phi_m-\phi)^2$ with the  maximum packing fraction $\phi_m \approx 0.67$.  
$\rho(\phi)$ is  the effective density given by $\rho(\phi)= (\rho_p \phi + \rho_f(1-\phi))$, where 
$\rho_f$ and $\rho_p$ are the densities of the fluid and particulate phases, respectively. 
The velocity $\u$ is the average of the two velocities of the fluid phase and the particle phase, 
and this average is defined as
\begin{eqnarray*}
\u = (1-\phi)\u_f + \phi\u_p  \hspace{0.5cm}\hbox{ in }\hspace{0.3cm}0 < y < h(x, t), 
\hspace{0.3cm}t\geq 0,  \nonumber 
\end{eqnarray*}
where $\u_f$ and $\u_p$ are the velocities of the fluid and particulate phases, respectively. 
At the free surface, the normal stress balance and the tangential stress balance are given by  
\begin{eqnarray}
\n\cdot \left(-p\I+\mu(\phi)(\nabla\u + \nabla\u^{T})\right)
\cdot \n&=& \gamma \kappa,        \label{BC2}\\
\t \cdot \left(-p\I+\mu(\phi)(\nabla\u + \nabla\u^{T})\right)
\cdot \n&=& 0,      
\hspace{0.5cm}\hbox{ at }\hspace{0.3cm}y = h(x, t),  \label{BC3} 
\end{eqnarray}
where $\n$ and $\t$ are the unit outward normal and unit tangential vectors to the free surface, 
$\gamma$ is the surface tension of the mixture and independent of particle concentration, and  
$\kappa$ is the curvature of the free surface. 
The boundary condition at the wall is
\begin{eqnarray}
\u &=& 0  \hspace{0.5cm}\hbox{ at }\hspace{0.3cm}y = 0. \label{BC4}
\end{eqnarray}
Note that the governing equations {\eqref{NS1}}--{\eqref{BC4}} assume that $\phi$ is independent of $y$, which  prohibits particles from settling to the substrate. 
Therefore, these governing equations cannot explain the settling behaviors that occur in the settled regime (a).
Cook {\cite{C}} has proposed an equilibrium model which is based on the balance of the hindered settling and shear-induced migration. 
This model shows good agreements with the experimental data from Zhou et al. {\cite{ZDBH}}, and captures the transition between the well-mixed regime (c) 
and the settled regime. 
In addition, there are some experimental results and derivation of models (see {\cite{MB, MHL, MPPB, WB, WM}}). 
\par 
Cook et al. {\cite{CBH}} has revisited a lubrication model of Zhou et al. {\cite{ZDBH}} with more complete explanations. 
The lubrication model mentioned there is derived from the governing equations {\eqref{NS1}}--{\eqref{BC4}} as the following  
system of conservation laws (see {\cite{CBH, ZDBH}}): 
\begin{eqnarray}
\begin{cases}
\displaystyle\frac{\partial}{\partial t}h + 
\frac{\partial}{\partial x}\left(h^3\frac{\rho(\phi)}{\mu(\phi)}\right) =0, \\
\\
\displaystyle\frac{\partial}{\partial t}(\phi h) + 
\frac{\partial}{\partial x}\left(h^3\phi\frac{\rho(\phi)}{\mu(\phi)}+v_s h \phi(1-\phi)F(\phi)W(h)\right) =0, 
\end{cases}
\label{system}
\end{eqnarray}
where $F(\phi) = (1-\phi)^{5.1}$ and $v_s=\dfrac{2}{3}\dfrac{(\rho_p-\rho_f)}{\rho_f}\left(\dfrac{a}{h_0}\right)^2$, 
$a$ is the distance from the center of the particle to the wall and 
\begin{eqnarray}  
W(h) = \dfrac{\frac{1}{18}(\frac{h}{a})^2}{\sqrt{1+[\frac{1}{18}(\frac{h}{a})^2]^2}}.  \label{wall} 
\end{eqnarray}
\noindent 
The term $v_s h \phi(1-\phi)F(\phi)W(h)$ represents the relative velocity of particles to the fluid incorporating with the interference among particles themselves by means of $F(\phi)$, which is known as {\em Richardson-Zaki correction} and is valid for high concentration of particles (cf. \cite{MPPB, RZ, ZDBH}), and among particles and the wall approximately given by $W(h)$ in (\ref{wall}).
The function $W(h)$ is introduced in {\cite{CBH, ZDBH}} and has the asymptotics
\begin{equation}
\label{wall-asymptotic} 
W(h) \approx \begin{cases}
0 & \text{$h < a$}, \\
1 & \text{$h \gg a$}.
\end{cases}
\end{equation}
In \cite{CBH, ZDBH}, solutions of {\eqref{system}} with initial jumps in the height from the upstream film thickness $h_L$ to the precursor thickness $b$ and given concentrations on tips and ends,
that is 
\begin{eqnarray}
h(x, 0) =  
\begin{cases}
h_L &\hbox{ for }x\leqq 0 \\
b &\hbox{ for }x>0
\end{cases}
, \hspace*{0.5cm}\phi(x, 0)= \begin{cases}
\phi_{0L} &\hbox{ for }x\leqq 0 \\
\phi_{0R} &\hbox{ for }x>0
\end{cases},  
\label{initial}
\end{eqnarray}
are treated {\em only in the asymptotic case $W(h) \approx 1$}, where $h_L$, $b$, $\phi_{0L}$ and $\phi_{0R}$ are particularly chosen constants.
For the initial value problem {\eqref{system}} and {\eqref{initial}}, namely the {\em Riemann problem} of  {\eqref{system}}, authors have found numerical solutions 
with a $1$-shock wave from the upstream state $(h_L, \phi_{0L} h_L)$ to an intermediate state $(h_I, \phi_I h_I)$ with height $h_I$ and concentration $\phi_I$  slightly larger than $h_L$ and $\phi_{0L}$, respectively,
and a $2$-shock wave from this intermediate state to the precursor $(b, \phi_{0R} b)$.  
According to their results, the double-shock wave ($1$-shock and $2$-shock) can generate the particle-rich ridge, which looks \lq\lq a bump", observed in the ridged regime (b). 
Despite good agreements with experiments and simple description of structures, the explicit dependence of several particle characteristics such as {\em radius} is not clearly mentioned these results, which can yield misunderstandings to interpret phenomena for concrete particle laden flows.
\par 
Our main aim here is to unravel one of these unclear points in detail.
In particular, the effect of particles with nontrivial radii on morphology of particle laden flows is explicitly considered, which includes the asymptotics (\ref{wall-asymptotic}) in the Riemann problem for {\eqref{system}} with the initial condition {\eqref{initial}}. 
The present study then lead us to reveal the genuine influence of interaction between nontrivial-size particles and the wall on the formation of the particle-rich ridge through the shock structure. 
The organization of this paper is as follows. 
In Section {\ref{sec:2}}, we briefly review a theory of shock waves for the Riemann problem of the system {\eqref{system}} with {\eqref{initial}}. 
In mathematical theory, it is known that the general $m\times m$ system of the hyperbolic conservation laws   
\begin{eqnarray*}
\partial_tU + \partial_x(F(U)) = 0  
\end{eqnarray*}
can have discontinuous solutions such as shock waves and smooth solutions such as rarefaction waves, 
where $U= (U_1,\cdots, U_m)^{\top}\in\R^m$, $(x,t)\in \R\times \R_{+}$, $\R_+\equiv [0, +\infty)$ 
and $F(U) = (F_1(U),\cdots, F_m(U))^{\top}$ is a vector-valued function which is $C^2$ in some open subset $D\subset \R^m$ 
(see {\cite{La1, S}}). In order to construct the solution with shock waves of the system {\eqref{system}}, 
we consider the case where the solutions have a discontinuity, and hence we deal with {\em weak solutions} mentioned below. 
Weak solutions consisting of shock waves are constructed applying mathematical theories established in {\cite{La1, S}} to the system {\eqref{system}}. 
In Section {\ref{sec:4}}, families of weak solutions of {\eqref{system}} are numerically studied.
In particular, we pay attention to the particle-rich ridge from the mathematical viewpoint.
To this end, the \lq\lq bump"-like solutions are constructed by means of physically relevant shock waves.
We especially investigate characteristics of the family of solutions with the required structure, such as the range of existence, height and propagation speed.
In the present model {\eqref{system}}, solutions can depend on the radius $a$ whose explicit contributions are overlooked and overestimated in previous studies such as \cite{CBH, ZDBH}, and the concentration $\phi$ of particles in flows.
The present study mathematically reveals the qualitative nature of particle-rich ridge which nontrivial particle features are taken into account.

\section{Preliminaries from the theory of conservation laws}
\label{sec:2}
Here we apply terminologies in mathematical theory of conservation laws to (\ref{system}) with their brief review.
Setting $n=\phi h$, the system {\eqref{system}} is reformulated to the system of conservation laws for $(h,n)$ as follows: 
\begin{eqnarray}
\begin{cases}
\partial_t h + \partial_x\left(h^3f \left(\dfrac{n}{h}\right)\right) 
=0, \\
\\
\partial_t n + \partial_x\left(h^2 n f\left(\dfrac{n}{h}\right)
+ h g \left(\dfrac{n}{h}\right)W(h)\right) 
=0, 
\end{cases}
\label{system1}
\end{eqnarray}
where 
\begin{eqnarray*}
f\left(\dfrac{n}{h}\right)&=&\left(1+\dfrac{(\rho_p-\rho_f)}{\rho_f}\left(\dfrac{n}{h}\right)\right)
\cdot\dfrac{(\phi_m-\phi)^2}{\phi_m^2}, \\
g\left(\dfrac{n}{h}\right)&=&\dfrac{2}{3}\dfrac{(\rho_p-\rho_f)}{\rho_f}\dfrac{a^2}{h_0^2}
\left(\dfrac{n}{h}\right)\phi(1-\phi)^{6.1}.
\end{eqnarray*} 
Letting
\begin{eqnarray*}
U = 
\begin{pmatrix}
h \\
n 
\end{pmatrix}
, 
\hspace{0.5cm}
F(U) =
\begin{pmatrix}
h^3f \left(\dfrac{n}{h}\right) \\
h^2 n f\left(\dfrac{n}{h}\right)+ h g \left(\dfrac{n}{h}\right)W(h)
\end{pmatrix}, 
\end{eqnarray*}
the system (\ref{system1}) can be rewritten in the form 
\begin{eqnarray}
\partial_t U + \partial_x(F(U)) = 0.  \label{conservation}
\end{eqnarray}
It is well known that a solution to conservation laws {\eqref{system}} can be 
discontinuous even if the initial profile is smooth. 
We therefore treat weak solutions which are defined as follows : 
\begin{definition}[\cite{S}]
A bounded measurable function $U(x, t)$ on $\R\times \R_{+}$ is called a {\em weak solution}
of the initial-value problem for {\eqref{conservation}} with bounded 
and measurable initial data $U(x, 0)$, provided that 
\begin{eqnarray}
\int^{\infty}_0 \int_{\R} (U\psi_t + F(U)\psi_x)dxdt 
+ \int_{\R} U(x, 0) \psi(x, 0) dx = 0 \label{weak}
\end{eqnarray}
holds for all $\psi \in C_0^1(\R\times\R_{+})$, where $C^1_0(X)$ is the space of continuously differentiable functions defined on a set $X$ with compact supports in $X$. 
\end{definition}
\noindent 
If the weak solution $U(x, t)$ has a discontinuity along a curve $x = x(t)$, 
the solution $U$ and the curve $x = x(t)$ must satisfy the 
{\em Rankine-Hugoniot relation} (jump condition)
\begin{eqnarray}
s(U_L-U_R) = F(U_L)-F(U_R),  \label{RH1}
\end{eqnarray}
where $U_L=U(x(t)\,-\,0, t)$ is the limit of $U$ approaching $(x,t)$ from the
left and $U_R=U(x(t)\,+\,0, t)$ is the limit of $U$ approaching $(x,t)$ from 
the right, and $s = \frac{dx}{dt}$ is the propagation speed of $x(t)$. 
The discontinuity is referred to as a {\em shock wave}, or simply a {\em shock}.
\par 
We consider the {\em Riemann problem} for the conservation laws {\eqref{system1}}, namely the initial value problem with the initial condition of the following form, which is called the {\em Riemann data}:
\begin{eqnarray}
U(x, 0)=
\begin{cases}
U_L \hspace{0.5cm}x<0 \\
U_R \hspace{0.5cm}x>0 
\end{cases}
. \label{initial1}
\end{eqnarray} 
The Jacobian matrix of $F$ at $U$ is 
\begin{eqnarray*}
&DF(U)= 
\begin{pmatrix}
F_{11}(U) & F_{12}(U) \\
F_{21}(U) & F_{22}(U)    
\end{pmatrix},
\end{eqnarray*}
where 
\begin{align*}
F_{11}(U) &= 3h^2f-n h(d_{\phi}f),\\ 
F_{12}(U) &= h^2(d_{\phi}f), \\
F_{21}(U) &= 2nhf-n^2(d_{\phi}f) + \left( -\displaystyle\frac{n}{h}(d_{\phi}g)+ g \right) W + hg (d_h W),\\ 
F_{22}(U) &= h^2f+n h (d_{\phi}f)+(d_{\phi}g)W,
\end{align*}
and their eigenvalues, which are also called {\em characteristic fields}, of $DF(U)$ are 
\begin{eqnarray}
\lambda_1 (U) = \displaystyle\frac{P-\displaystyle\sqrt{Q}}{2},  
\hspace{0.5cm}
\lambda_2 (U) = \displaystyle\frac{P+\displaystyle\sqrt{Q}}{2}, \label{eigen}    
\end{eqnarray}
where
\begin{equation*}
P= F_{11}(U) + F_{22}(U),\quad Q = \{F_{11}(U) - F_{22}(U)\}^2 + 4F_{12}(U)F_{21}(U).
\end{equation*}
If we assume $Q>0$, eigenvalues $\lambda_j(U)$ $(j=1,2)$ are real-valued and $\lambda_1(U)<\lambda_2(U)$ 
holds for any $U\in\Omega$, where $\Omega = \{(h,n)\in\R^2 : 0<h,\,\,0\leq n< \phi_m h\}$. 
In other words, the system {\eqref{system1}} is {\em strictly hyperbolic} in $\Omega$.  
The right eigenvectors associated with the eigenvalues $\lambda_j(U)$ are 
\begin{eqnarray*}
r_1(U) = 
\begin{pmatrix}
-2F_{12}(U)  \\
Z+ \displaystyle\sqrt{Q}     
\end{pmatrix}
,\hspace{0.5cm}
r_2(U) = 
\begin{pmatrix}
-2F_{12}(U)  \\
Z- \displaystyle\sqrt{Q}     
\end{pmatrix}, 
\end{eqnarray*}
where $Z= F_{11}(U) - F_{22}(U)$. 
To characterize characteristic fields further, we calculate the quantities
\begin{equation*}
\nabla \lambda_j(U) \cdot r_j(U),\quad j=1,2
\end{equation*}
for each $U\in \Omega$. 
If $\nabla \lambda_j(U) \cdot r_j(U) \not = 0$ holds in a subset $N\subset \Omega$, the $j$-th characteristic field $\lambda_j(U)$ is called {\em genuinely nonlinear} in $N$, which ensures the existence of solutions of the Riemann problem with the left state $U_L$ in $N$ and right states close to $U_L$.
We omit the detailed calculations of the quantities here since lengthy calculations are required. 
Instead, we numerically confirm later that both characteristic fields are genuinely nonlinear in regions of our interest.
In this case, weak solutions of the Riemann problem consist of at most three constant states 
$U_L$, $U_I$, $U_R$ followed by either shock waves or rarefaction waves (e.g. {\cite{La1, S}}).
\par
In the present argument, our interest is restricted to solutions to (\ref{system1}) consisting of discontinuities, namely shock profiles.
As mentioned, discontinuities arising in weak solutions of (\ref{conservation}) have constraints determined by the Rankine-Hugoniot relation \eqref{RH1}.
To calculate such discontinuities for (\ref{system1}), we fix the reference point $U_L=(h_L, n_L)$ and consider right states $U_R=U=(h, n)$ solving \eqref{RH1}.
If a weak solution of (\ref{system1}) has a jump discontinuity between the left state
$U_L$ and the right state $U$, then $U$ must satisfy 
\begin{eqnarray}
\begin{cases}
s(h-h_L) = F_1(h, n) - F_1(h_L, n_L), \\
s(n-n_L) = F_2(h, n) - F_2(h_L, n_L), 
\end{cases}
\label{RH2}
\end{eqnarray}
where 
\begin{eqnarray*}
F_1(h, n)&=& h^3f\left(\dfrac{n}{h}\right), \\
F_2(h, n)&=& h^2 n f\left(\dfrac{n}{h}\right) + h g\left (\dfrac{n}{h}\right)W(h). 
\end{eqnarray*}
Eliminating $s$ from these equations, we obtain 
\begin{eqnarray}
\hspace{-0.4cm}
\dfrac{n - n_L}{h-h_L}-\dfrac{F_2(h, n)-F_2(h_L, n_L)}{F_1(h, n)-F_1(h_L, n_L)}=0, 
 \label{RH3}
\end{eqnarray}
whose graph is called the {\em Hugoniot locus}. 
Among solution families in the Hugoniot locus, physically relevant discontinuities additionally require the following $k$-{\em entropy inequalities} $(k=1,2)$ (e.g. \cite{S})   
\begin{eqnarray}
&&s<\lambda_1(U_L), \hspace{0.5cm}\lambda_1(U)< s< \lambda_2(U), 
\hspace{0.5cm} \text{($1$-entropy inequality)},
\label{1entropy} \\
&& \nonumber \\
&&\lambda_1(U_L)< s< \lambda_2(U_L),\hspace{0.5cm}\lambda_2(U)<s, 
\hspace{0.5cm} \text{($2$-entropy inequality)},  
\label{2entropy}
\end{eqnarray}
where $s$ is the speed of discontinuities (shocks):
\begin{eqnarray*}
s=\dfrac{F_1(h, n)-F_1(h_L, n_L)}{h-h_L}=\dfrac{F_2(h, n) - F_2(h_L, n_L)}{(n-n_L)}. 
\end{eqnarray*}
If $U$ satisfies {\eqref{RH3}} and {\eqref{1entropy}}, then 
$U$ can be connected to $U_L$ from the right followed by a {\em 1-shock wave}.
Similarly, $U$ can be connected to 
$U_L$ from the right followed by a {\em 2-shock wave}, 
provided $U$ satisfies {\eqref{RH3}} and {\eqref{2entropy}}.    

\section{Dependence of radius and concentration of particles on bump generation}
\label{sec:4}

\begin{figure}[htbp]\em
	\begin{minipage}{1.0\hsize}
		\centering
		\includegraphics[width=8.0cm]{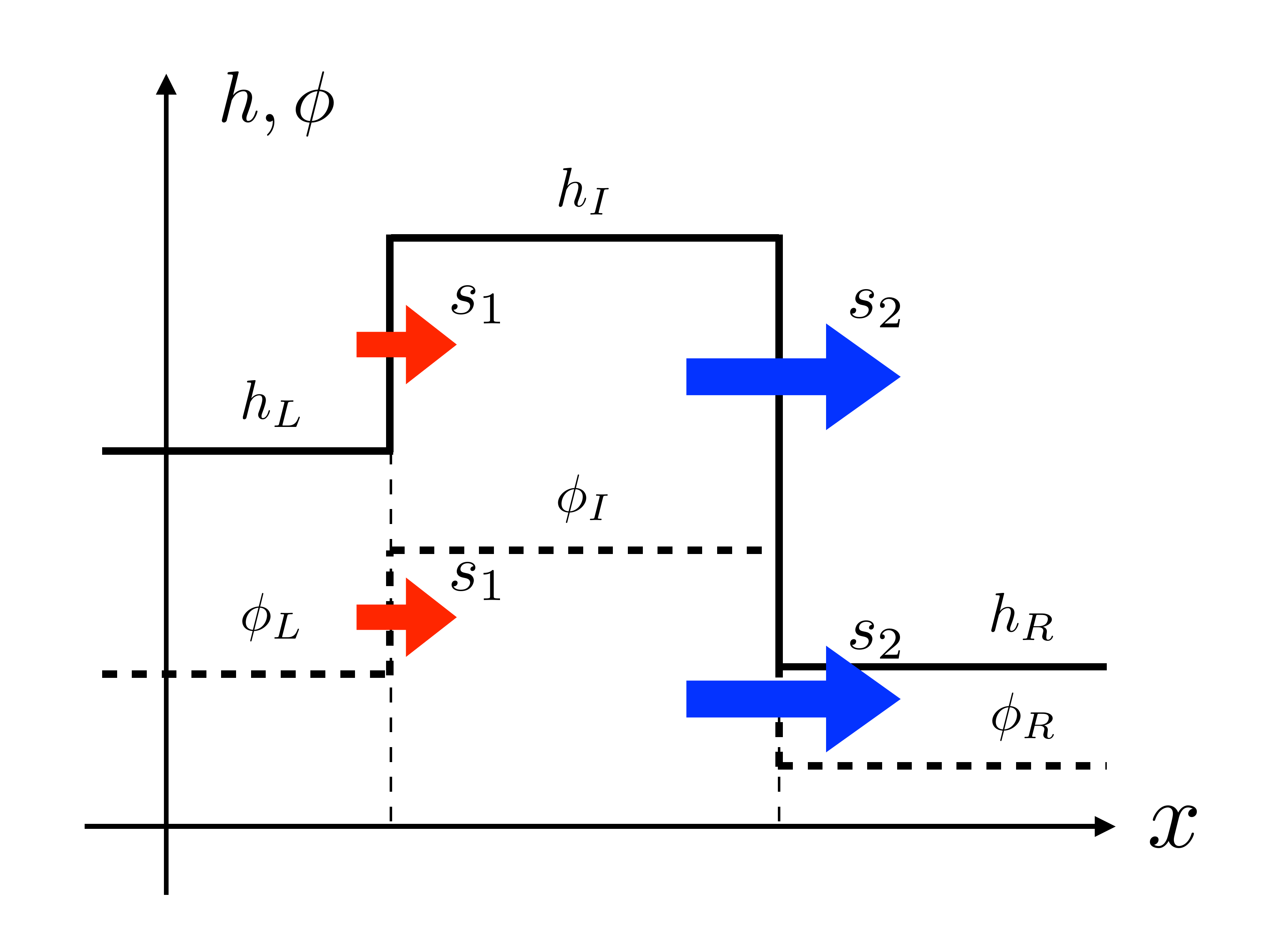}
	\end{minipage}
	\caption{The solution structure with bump generated by Lax's shock waves}
	\flushleft
	Solid lines describe the $h$-component of solutions for the Riemann problem \eqref{system}, while the dotted (thick) lines describe the $\phi$-component.
	The higher values of the intermediate state $(h_I, \phi_I)$ compared with both states given as the Riemann data $(h_L, \phi_L), (h_R, \phi_R)$ show the generation of bump structures.
	Two shocks with the speeds $s_1$ and $s_2$, respectively, do not interact during time evolution as long as $s_1 < s_2$.
	\label{fig-shock_bump}
\end{figure}

Our main concern is the investigation of \lq\lq bump" structure described by the composite waves of shocks.
In particular, we study the existence of solutions of the Riemann problem {\eqref{system1}}, {\eqref{initial}} consisting of three states $\{(h_L, \phi_L), (h_I, \phi_I), (h_R, \phi_R)\}$ such that
\begin{description}
\item[(B1)] The leftmost state $(h_L, \phi_L)$ is connected to an intermediate state $(h_I, \phi_I)$ with $h_L < h_I$ and $\phi_L < \phi_I$ followed by a $1$-shock with the shock speed $s_1$.
\item[(B2)] The state $(h_I, \phi_I)$ is connected to the rightmost state $(h_R, \phi_R)$ with $h_R < h_I$ and $\phi_R < \phi_I$ followed by a $2$-shock with the shock speed $s_2 > s_1$.
\end{description}
A schematic illustration, namely the physical realization, of weak solutions of the above form is shown in Figure \ref{fig-shock_bump}.
As mentioned in Introduction, $h_L$ represents the upstream film flow thickness, while $h_R$ denotes the precursor thickness\footnote{
Distributed flows from the tip of bumps whose height can be lower than $h_R$ is not considered in the present model, which will be studied through the general initial value problem of (\ref{system}), or the unreduced model (\ref{NS1}).
}.
The requirement for distributions of three individual states claims the existence of bump structure\footnote{
The bump structure in solutions to \eqref{system} in the present argument means a pair of step-up and step-down functions, as described in Figure \ref{fig-shock_bump}.
} in the particle laden generated by shock waves.
The additional requirement for shock speeds indicates that the bump structure persists in the sense that two shocks do not interact during time evolution, otherwise two shocks collide with each other to create a single shock structure (cf. \cite{S}).
We shall call the above structure a {\em bump structure generated by Lax's shocks} for simplicity.
We do not require any relationships between $h_L$ and $h_R$, and $\phi_L$ and $\phi_R$, respectively. 
\par
Analytic treatments of the present problem with concrete data are not easy tasks due to the complexity of nonlinearity and eigensystems $\{\lambda_j(U), r_j(U)\}_{j=1,2}$.
We instead study the structure of weak solutions, in particular shocks, numerically.
Our objective here is then reduced to numerical computations of Hugoniot loci, namely solutions of \eqref{RH2}, through the fixed left states $(h_L, \phi_L)$, and possibly intermediate states $(h_I, \phi_I)$.
As mentioned in the previous section, the pair of functions $(h,n) = (h,\phi h)$ instead of $(h,\phi)$ itself is considered so that the mathematical theory of conservation laws is directly applied.
In \eqref{RH2} there are three unknowns $(h,n,s)$, whereas the system \eqref{system} consists of two equations.
In order to ensure the unique solvability, we regard one of unknowns as a parameter and compute the loci as $1$-parameter families of solutions.
In the present study, we regard $h$ as a parameter and the remaining unknowns $(s, n)$ as functions of $h$, which induce the following parameter-dependent zero-finding problem:
\begin{equation}
\label{zero-finding}
G({\bf x};  {\bf p}) = 0,\quad G:\mathbb{R}^2\times \mathbb{R}^k\to \mathbb{R}^2,\quad {\bf x} = (s,n)^T,\quad {\bf p} = (p,\mu)^T,
\end{equation}
where ${\bf x}$ is the phase variable vector and ${\bf p} = (p,\mu)$ is the parameter vector consisting of varying parameter $p\in \mathbb{R}$ and the remaining fixed parameter vector $\mu \in \mathbb{R}^{k-1}$.
In the present case, we set 
\begin{equation*}
p = h,\quad \mu = (h_p, n_p, a, \rho_p, \rho_f, \phi_m),
\end{equation*}
The latter represents a given left state $(h_p, n_p)$ and the environment parameters of the system.
The nonlinearity $G$ becomes
\begin{equation}
\label{zero-conservation}
G({\bf x}; {\bf p}) = \begin{pmatrix}
h^3 f\left(\frac{n}{h}\right) - h_p^3 f\left(\frac{n_p}{h_p}\right) - s(h-h_p) \\
\left\{h^2 n f\left(\frac{n}{h}\right) - h_p^2 n_p f\left(\frac{n_p}{h_p}\right)\right\} + \left\{h g\left(\frac{n}{h}\right) - h_p g\left(\frac{n_p}{h_p}\right)\right\} - s(n-n_p)
\end{pmatrix}.
\end{equation}
We then apply the {\em pseudo-arclength continuation method} well-established in numerical bifurcation theory (e.g. \cite{DKK, K}) to (\ref{zero-finding}) for obtaining Hugoniot loci connected to $(h_p, n_p)$.
In this methodology, the parameter $h$ is regarded as the additional variable.
From the definition, the triple $\{(h_p, n_p), (h, n), s\}$ solving (\ref{zero-finding}) generates a shock with speed $s$.
\par
To ensure that the triple $\{(h_p, n_p), (h, n), s\}$ generates a shock in the sense of Lax, we verify the entropy conditions \eqref{1entropy} and \eqref{2entropy}. 
If either \eqref{1entropy} or \eqref{2entropy} is satisfied, we can conclude that the triple $\{(h_p, n_p), (h, n), s\}$ generates a $1$-shock if \eqref{1entropy} is satisfied, while it generates a $2$-shock if \eqref{2entropy} is satisfied, as long as the original system \eqref{system} satisfies the {\em strict hyperbolicity} in the domain including the chain of loci of our interest.
See Remark \ref{rem-hyp} below about the strict hyperbolicity of the present problem. 
\par
\bigskip
Our operations mentioned here are summarized as follows.
\begin{enumerate}
\item Fix $(h_p, n_p) = (h_L, n_L)$ and solve (\ref{zero-finding}).
\item For each solution $(h,n,s)$, verify the entropy conditions \eqref{1entropy} and \eqref{2entropy}.
\item Fix a point $(h, n)$ on the obtained loci as a new left state $(h_p, n_p)$, which corresponds to an intermediate state $(h_I, n_I)$, and go back to 1. 
\end{enumerate}
\par
Following our purpose, we restrict the operation 3 to the case that $\{(h_p, n_p), (h, n), s\}$ satisfies the $1$-entropy condition \eqref{1entropy} and the first our requirement between $(h_L, n_L)$ and $(h, n)$, namely $h_L < h$ and $\phi_L \equiv n_L / h_L < n/h \equiv \phi$.

\begin{remark}
When we compute Hugoniot loci for the system of conservation laws, typical unknown variables corresponding to ${\bf x}$ in (\ref{zero-finding}) are the state variables $(h,n) \equiv U$ in \eqref{RH2} and the shock speed $s$ is regarded as a component of parameters.
However, $s$ is determined by the left state $(h_p, n_p)$ and the right state $(h, n)$ and is not easy to predict a priori.
One can eliminate $s$ through the Rankine-Hugoniot condition like (\ref{RH3}), which induces extra calculations for constructing the corresponding zero-finding problem and unnecessary complexity of the system.
On the other hand, the dependence of states $(h,n)$ in our interest around $(h_L, n_L)$ is very clear. 
For example, the initial guess of $h$ should be sufficiently close to $h_L$.
Therefore it is easy to control solutions $(n, s)$ depending on the \lq\lq parameter" $h$, which yields the present methodology.
\end{remark}

\begin{remark}
\label{rem-right}
In typical (and mathematical) treatments of the Riemann problem, in addition to the left state, the right state $(h_R, n_R)$ is also fixed as the initial state (cf. \cite{CBH}).
On the other hand, the right state is regarded as one of unknowns in (\ref{zero-finding}) because our interest here is the possible distribution of states generating shock structures of our interests, while the Riemann problem with {\em given end states} as the initial one focuses on all possibilities of solution structures including rarefactions.
\end{remark}

In the present computations, the strict hyperbolicity of the system \eqref{system} and the {\em genuine nonlinearity} for characteristic fields $\{\lambda_1(U), \lambda_2(U)\}$ are also calculated independently.
Note again that former requires to separate the solution structure among different type of elementary waves including shocks of Lax's type, and the latter ensures the mathematical rigor to the existence of shock chains, at least for the right state $(h_R, n_R)$ sufficiently close to the left state $(h_L, n_L)$.

\par
\bigskip
Here we pay attention to the shock structure of the system with particles under {\em negative buoyancy}.
The corresponding situation determines the parameter $\rho_p$.
Unless otherwise noted, we fix $\rho_p = 2500$ for arguments in this section, while the remaining parameter values $(\rho_f, \phi_m) = (970, 0.67)$ are fixed in the whole computations\footnote{
These values are quoted from Table I in \cite{WB}.
}.
The parameter $\phi_m$ represents the {\em maximal} concentration of particles in flow, which is required for all solutions of (\ref{system}).
Fortunately, all solutions we have computed and shown below satisfy $\phi < \phi_m$.
\par
Finally, we should mention the treatment of the ratio $(a/h_0)^2$ appeared below (\ref{system}), which is one of the main interests in the present study. 
In previous works such as \cite{C, ZDBH}, the ratio $(a/h_0)^2$ is not clearly described\footnote{
It can relate to the sensitive dependence of experimental data on precursor film, according to \cite{ZDBH}.
}.
On the other hand, $(a/h_0)^2$ is set around $0.01$ in following works, e.g. \cite{MPPB}.
At present we can choose the ratio $(a/h_0)^2$ arbitrarily since our arguments are based on purely mathematical ones, and hence the solution structure for several sample values of $(a/h_0)^2$ is available to see the qualitative similarity and difference.

\begin{remark}
\label{rem-hyp}
As shown in figures mentioned below, strict hyperbolicity and genuine nonlinearity are satisfied in the domain of $(h, \phi)$ where the shock structure is considered, as far as we have numerically verified.
Therefore we do not care about the failure of these properties during the following arguments.
\end{remark}

First we review the actual distribution of the function $W(h)$ in (\ref{wall}).
The sample values $W(1) \equiv W(h)|_{h=1}$ itself and its derivative for fixed values of the particle radius parameter $a$ are shown in Table \ref{Tab:W}.
Indeed, for sufficiently small $a$ such as $0.01$, the asymptotic behavior (\ref{wall-asymptotic}) is valid, whereas the differential $W'(1) \equiv (dW/dh)(1)$ provides a nontrivial contribution to various functions involving the system even for relatively small $a$.

\begin{table}[ht]
\centering
{
\begin{tabular}{ccc}
\hline
 $a$ & $W(1)$ & $W'(1)$\\
\hline\\[-3mm]
$0.01$ & $0.999998$ & $6.51247\times 10^{-6}$\\
$0.02$ & $0.999974$ & $1.04192\times 10^{-4}$\\
$0.05$ & $0.998985$ & $4.05793\times 10^{-3}$\\
$0.1$ & $0.998985$ & $6.20717\times 10^{-2}$\\
$0.2$ & $0.810979$ & $0.555772$\\
$0.5$ & $0.216517$ & $0.413147$\\
$0.99$ & $0.0564799$ & $0.112712$\\
$0.999$ & $0.05547$ & $0.110709$\\
$1.1$ & $0.0460486$ & $0.0917185$\\
\hline 
\end{tabular}%
}
\caption{The function $W(1)$ with fixed $a$}
\flushleft
\label{Tab:W}
\end{table}%

\subsection{Effect of particle radius variation on shock structure}
\begin{figure}[htbp]\em
	\begin{minipage}{0.33\hsize}
		\centering
		\includegraphics[width=7.0cm]{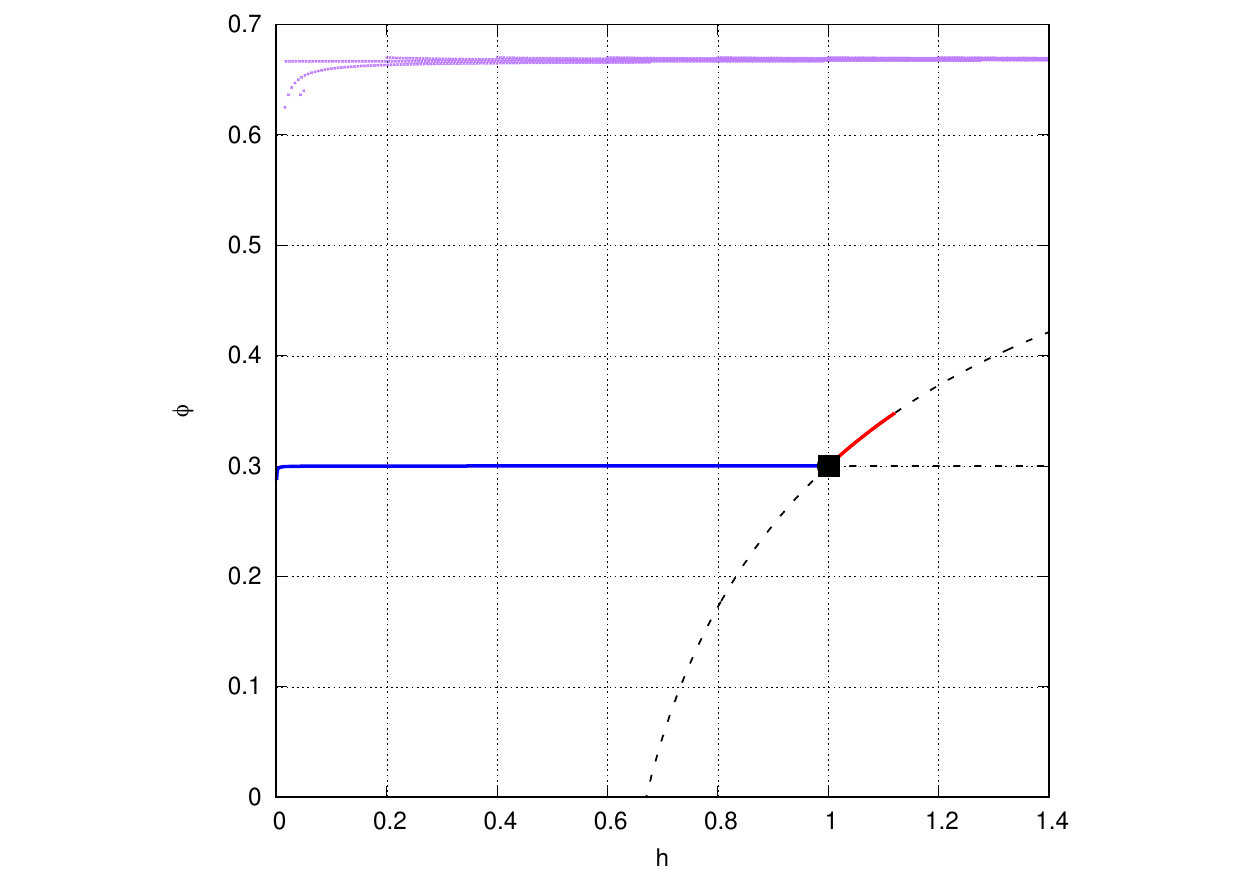}
	(a)
	\end{minipage}
	\begin{minipage}{0.33\hsize}
		\centering
		\includegraphics[width=7.0cm]{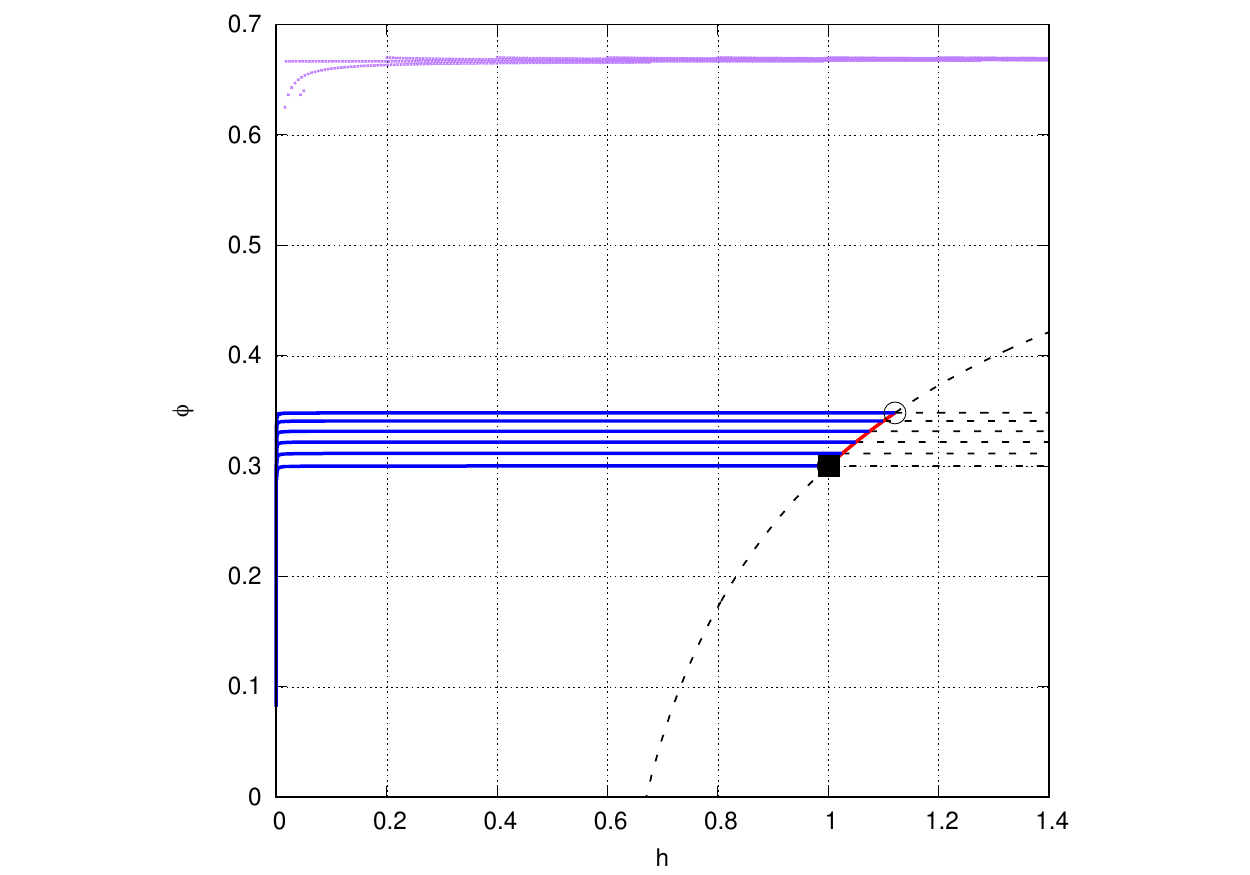}
	(b)
	\end{minipage}
	\begin{minipage}{0.33\hsize}
		\centering
		\includegraphics[width=7.0cm]{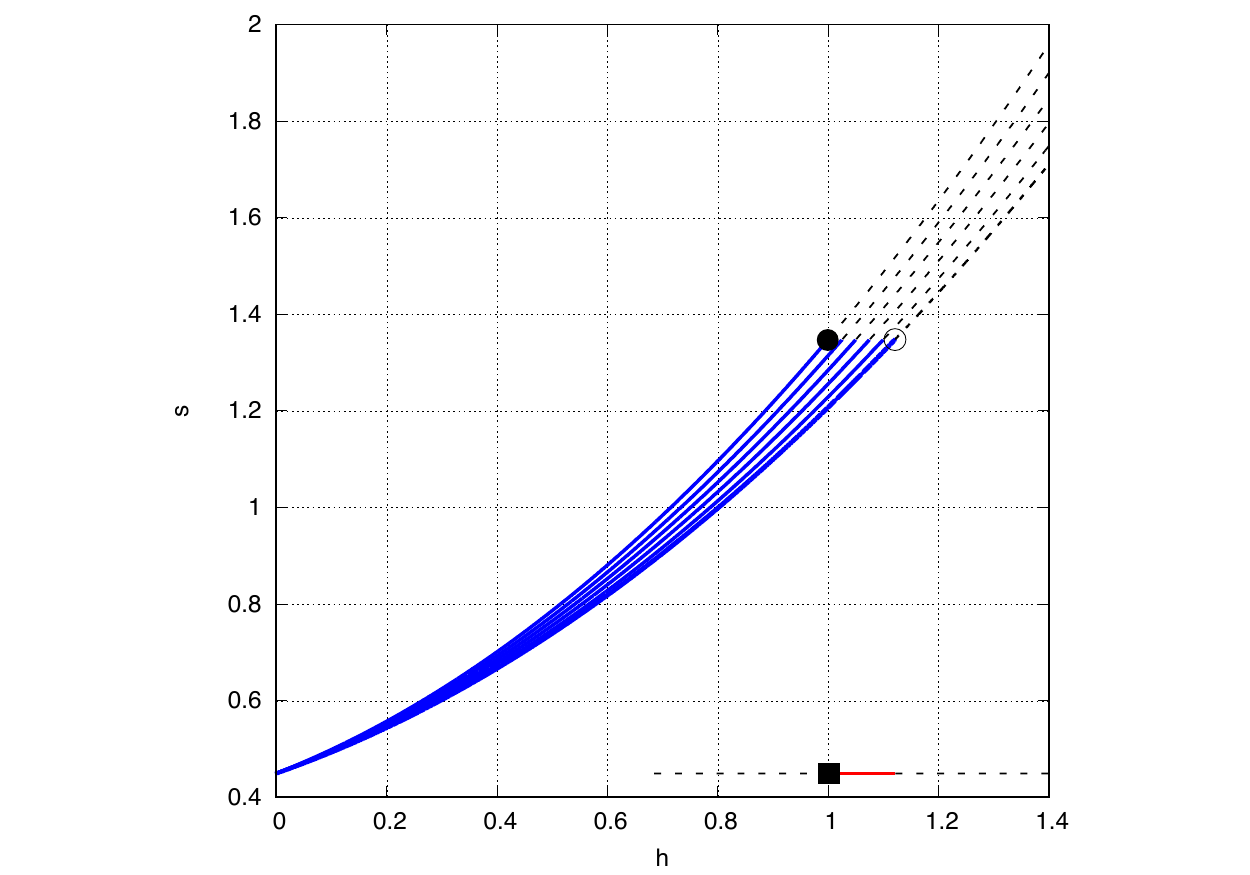}
	(c)
	\end{minipage}
	\caption{Solution families of (\ref{zero-finding}) with $(h_L, \phi_L) = (1.0, 0.3)$, $a=0.01$}
	\flushleft
	\label{fig-locus_sample}
	In all figures, the horizontal axis represents $h$ and the vertical axis represents $\phi$ in (a), (b), while the latter represents $s$ in (c). 
	The red curve denotes the collection of right states connected from the left states followed by $1$-shocks.
	The blue curve denotes the collection of right states connected from the left states followed by $2$-shocks.
	The dotted black curve denotes the collection of right states where the neither entropy conditions (\ref{1entropy}) nor (\ref{2entropy}) are satisfied.
	Finally, the purple region in (a) and (b) denotes the union of points where the genuine nonlinearity does not hold.
	\par
	The black square denotes the limit right state $(h,\phi)$ as $(h,\phi)\to (h_L,\phi_L)$ along the $1$-shock curve, while the black ball denotes the limit right state $(h,\phi)$ as $(h,\phi)\to (h_L,\phi_L)$ along the $2$-shock curve.
	The white ball denotes the limit right state $(h,\phi)$ as $(h,\phi)\to (h_I,\phi_I)$ along the $2$-shock curve, where $(h_I,\phi_I)$ is a point on the locus through $(h_L,\phi_L)$ followed by a $1$-shock.
	Here the boundary point of $1$-shock branch is drawn as the white ball in (b). 
	\par
	(a) shows the Hugoniot loci through $(h_L, \phi_L)$. 
	(b) shows the collection of Hugoniot loci through points on the $1$-shock curve (red).
	(c) shows the corresponding plot of shock speeds determined by right states drawn in (b). 
\end{figure}

Next we study the effect of $a$ on the distribution of shock structure.
To this end, we fix the left state as representatives
\begin{equation*}
h_0 = h_L = 1,\quad \phi_L = 0.3
\end{equation*}
and solve (\ref{zero-finding}).
Examples of Hugoniot loci representing right states centered at the left state $(h_L, \phi_L)$ followed by shocks are drawn in Figure \ref{fig-locus_sample}.
We typically find two branches of Hugoniot loci through $(h_L, \phi_L)$. 
One branch consists of right states followed by $1$-shocks in the direction where $h$ is increasing and pieces of states which Lax's conditions are not satisfied, while another branch consists of segments of right states followed by $2$-shocks in the direction where $h$ is decreasing and segments of states which Lax's conditions are not satisfied.
When right states admitting $1$-shocks are distributed in the region $\{h > h_L, \phi > \phi_L\}$, Hugoniot loci regarding a point on the locus in $\{h > h_L, \phi > \phi_L\}$ as the new left state $(h_I, \phi_I)$ are additionally computed (Figure \ref{fig-locus_sample}-(b)). 
From our purpose, only the loci admitting $2$-shocks are computed there. 
We see that the loci through various $(h_I, \phi_I)$ has the similar structure to that through $(h_L, \phi_L)$.
Our computations also show (in Figures \ref{fig-locus_sample}-(c)) that there are collections of right states $(h,\phi)$ connected from $(h_I,\phi_I)$ followed by $2$-shocks whose shock speeds $s_2$ are much higher than the speeds $s_1$ of the $1$-shocks connecting $(h_L,\phi_L)$ and $(h_I,\phi_I)$.
Consequently, the triple
\begin{equation*}
\{(h_L, \phi_L), (h_I, \phi_I), (h_R, \phi_R)\}
\end{equation*}
is a candidate of states generating the bump structure generated by Lax's shocks, where $(h_L, \phi_L)$ is fixed, $(h_I, \phi_I)$ is on the locus through $(h_L, \phi_L)$ followed by a $1$-shock satisfying $h_I > h_L$ and $\phi_I > \phi_L$, and $(h_R, \phi_R)$ is on the locus through $(h_I, \phi_I)$ followed by a $2$-shock (cf. Figure \ref{fig-shock_bump}).

\begin{figure}[htbp]\em
	\begin{minipage}{0.5\hsize}
		\centering
		\includegraphics[width=7.0cm]{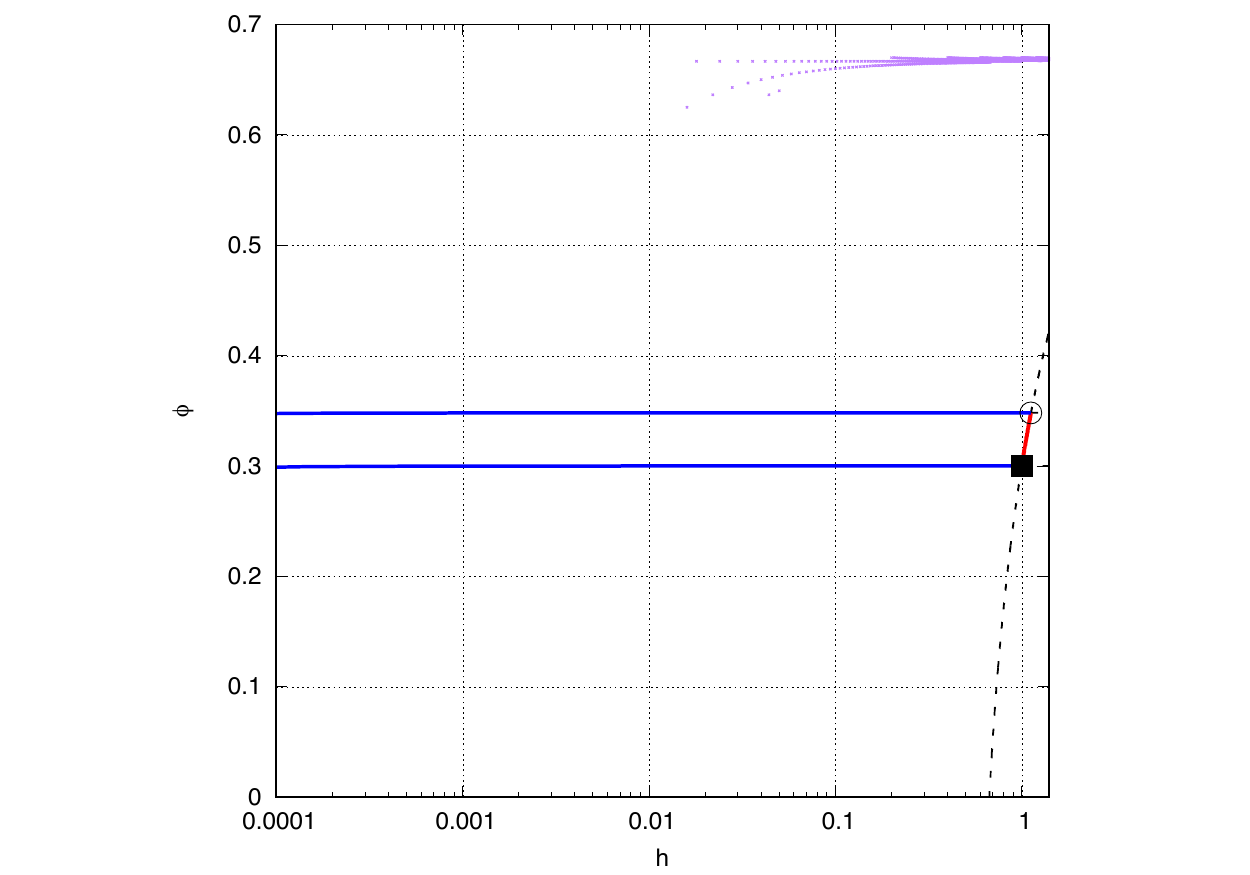}
	(a)
	\end{minipage}
	\begin{minipage}{0.5\hsize}
		\centering
		\includegraphics[width=7.0cm]{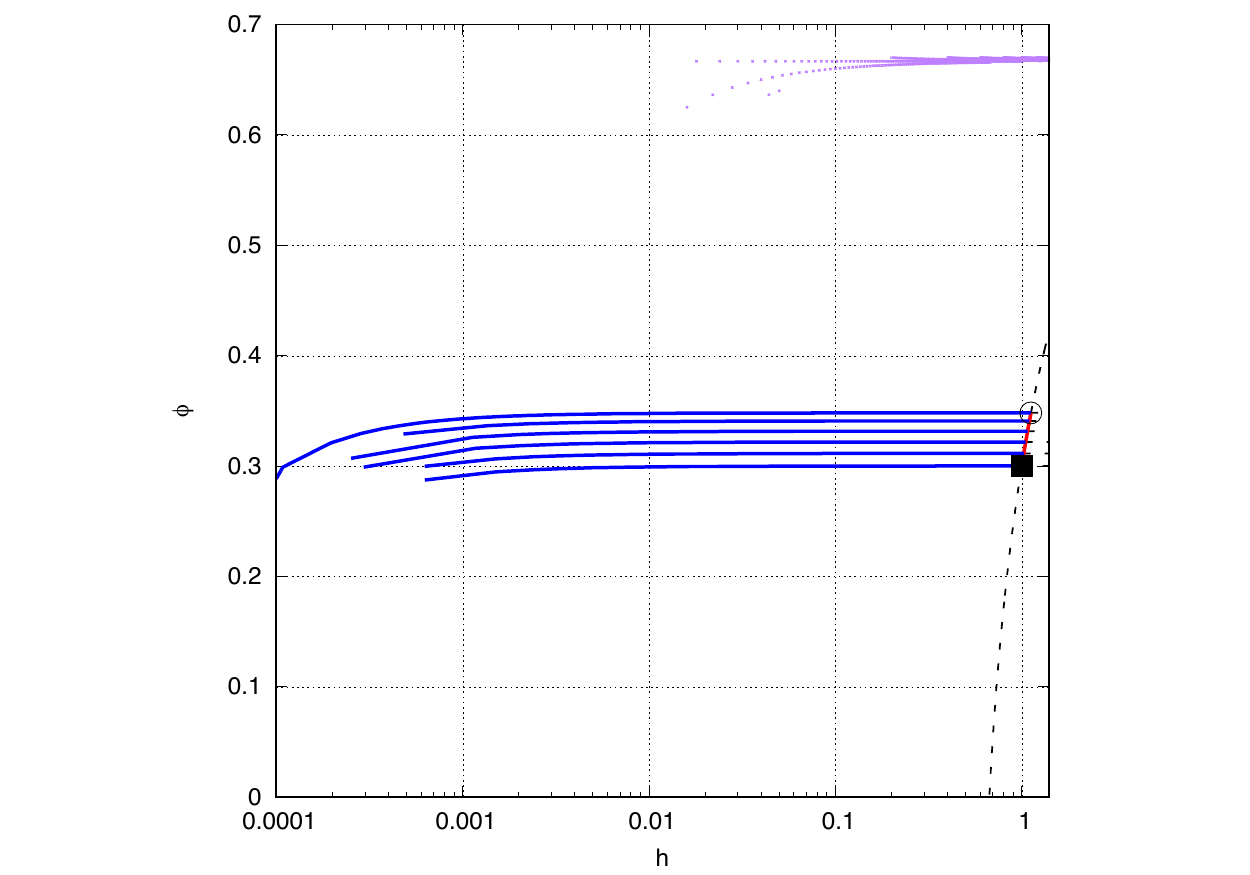}
	(b)
	\end{minipage}\\
	\begin{minipage}{0.5\hsize}
		\centering
		\includegraphics[width=7.0cm]{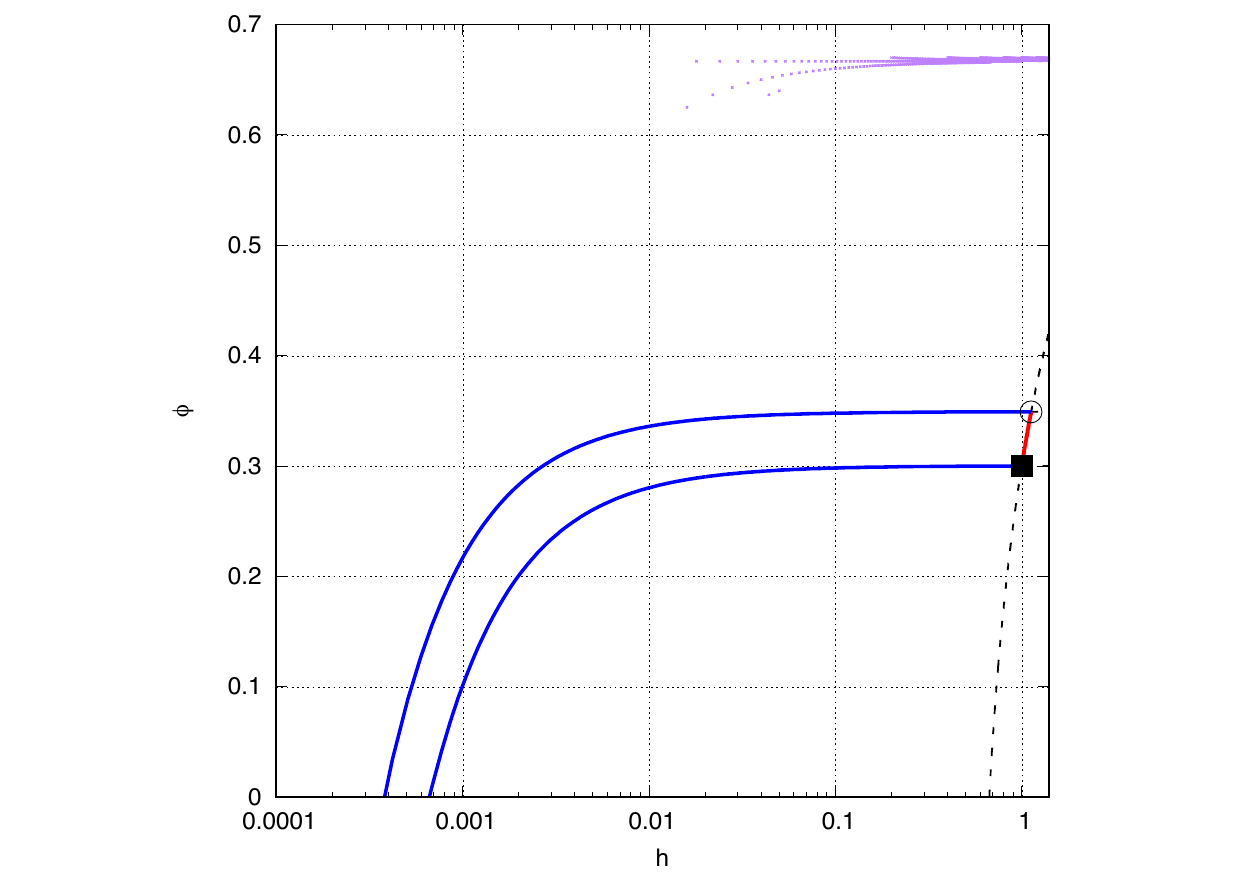}
	(c)
	\end{minipage}
	\begin{minipage}{0.5\hsize}
		\centering
		\includegraphics[width=7.0cm]{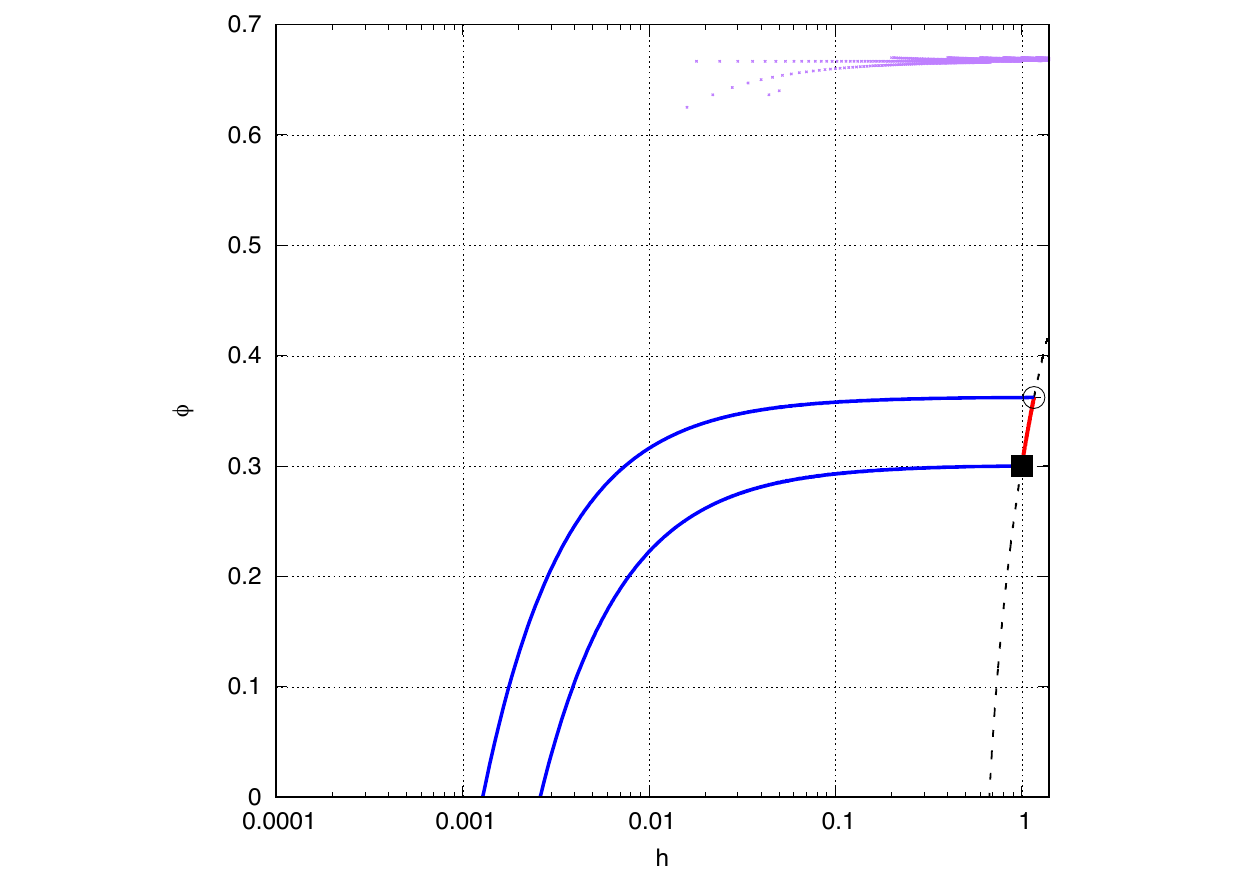}
	(d)
	\end{minipage}\\
	\begin{minipage}{0.5\hsize}
		\centering
		\includegraphics[width=7.0cm]{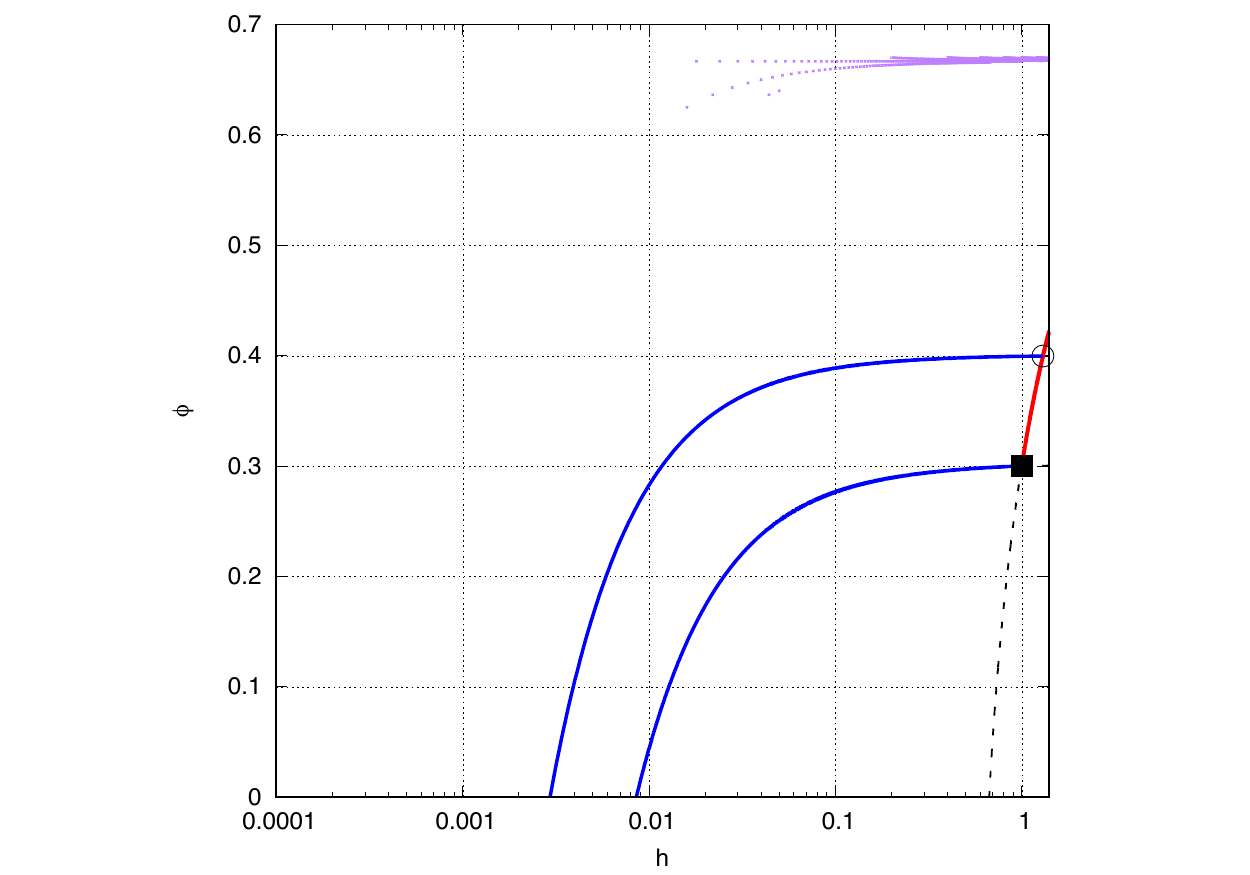}
	(e)
	\end{minipage}
	\begin{minipage}{0.5\hsize}
		\centering
		\includegraphics[width=7.0cm]{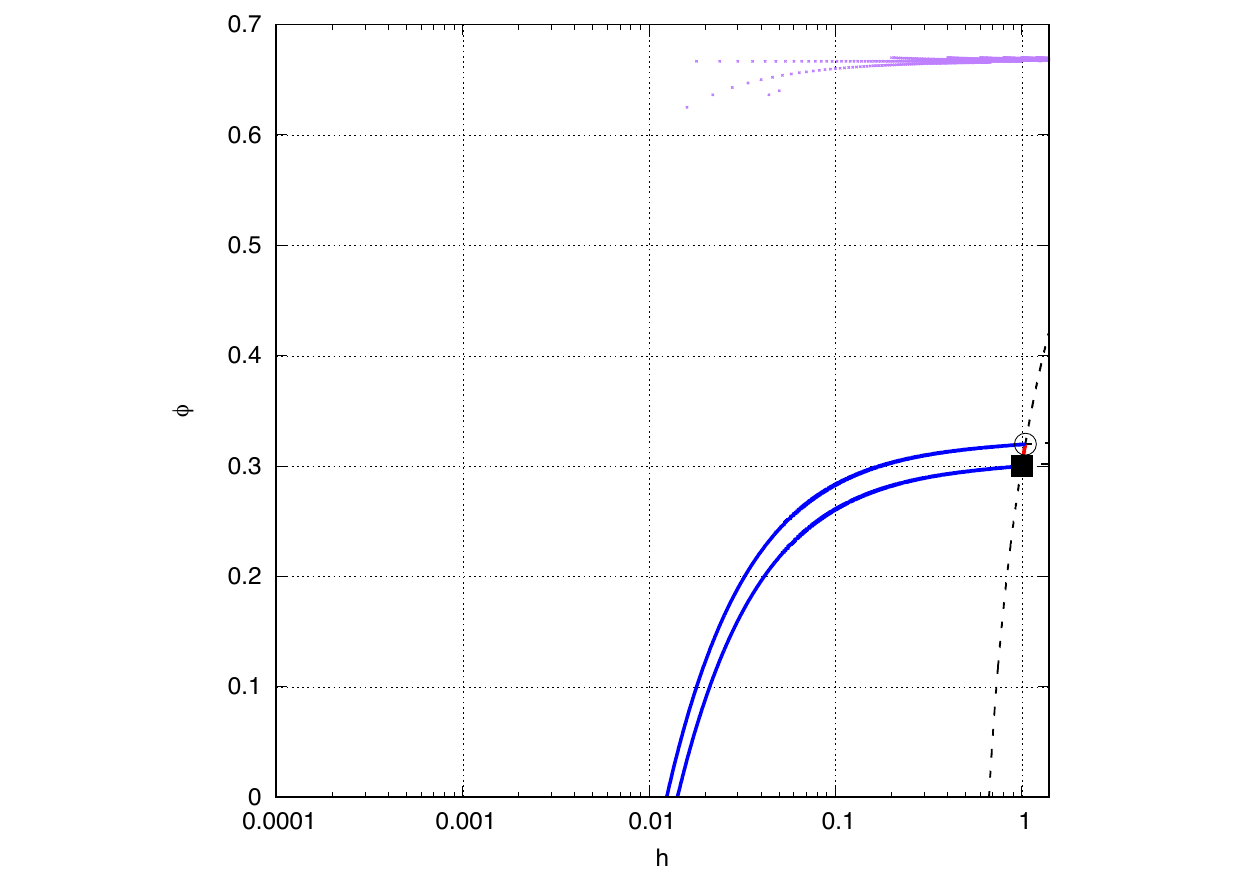}
	(f)
	\end{minipage}\\
	\caption{Solution families of (\ref{zero-finding}) with $(h_L, \phi_L) = (1.0, 0.3)$ and various $a$}
	\flushleft
	\label{fig-locus_var_a}
	The basic description rule of figures is the same as previous figures.
	In all figures, the horizontal axis is $h$ with the $\log$-scale plotting, while the vertical represents $\phi$. (a): $a=0.001$. (b): $a=0.01$. 
(c): $a=0.05$. (d): $a=0.1$. 
(e): $a=0.2$. (f): $a=0.5$. 
The length of $1$-shock curves depend on $a$.
Moreover, we see that, for small $a$ such as $a \leq 0.1$, the $2$-shock curves are almost flat for $h > a$.
\end{figure}

Changing $a$ during computation of Hugoniot loci with properties of our requirements, we observe (at least) two interesting features, which can be seen in Figure \ref{fig-locus_var_a}.
\begin{itemize}
\item The $\phi$-component of the right states followed by $2$-shocks is almost constant, namely the corresponding right states create plateau regions, as far as the ratio $a/h_R$ is approximately less than $1$. 
More precisely, $\phi$ behaves as a {\em monotonously increasing} function of $h$ on the plateau region.
\item For different choice of $a$, the length of curves of right states followed by $1$-shocks changes nontrivially. 
At least, the length does not change in the monotonous manner.
\item The speeds of $1$-shocks $s_1$ and $2$-shocks $s_2$ always satisfy $s_1 < s_2$ for all mentioned states observed in Figure \ref{fig-locus_var_a}.
In particular, our requirement about the speeds of shocks stated in (B2) is always satisfied for bumps generated by Lax's shocks, as far as we have computed.
\end{itemize}

The first observation reflects the validity of the model from the physical viewpoint.
The ratio $a/h$ measures the physical characteristic of films in finite-volume flows.
In order that particle laden flows behave like colloid, in particular continua, $(a/h_0)^2 \ll 1$ is required (cf. \cite{MPPB}\footnote{
It is noted as a reference that, $(a/h_0)^2 = O(10^{-2})$ in the end of early-stage transient evolution of particle laden flows, namely a well-mixed state which is a physically different situation from the present argument.
}).
Interestingly, the assumption $(a/h_0)^2 \ll 1$ approximately corresponds to the existence of plateau regions on $2$-shock curves as far as $a/h < 1$ holds for the right state.
In other words, we observe that loci of states in $(h,\phi)$-plane admitting $2$-shocks are strongly curved  if $a/h \geq 1$ holds or $a$ is relatively large, which can correspond to violation of continuum hypothesis.
Taking the treatment of particle laden flows as continua into account, which is essential to derive the present model, the plateau regions will provide physical reliability of arguments.
Moreover, the property of $\phi$ on the locus admitting $2$-shocks being an increasing function of $h$ is an important property for the following arguments.
\par
The second observation strongly relates to generations of the bump structure by Lax's shocks in various choice of particles in mixture fluids.
Indeed, Figure \ref{fig-locus_var_a} shows the collection of composite Hugoniot loci, where we can choose triples $\{(h_L, \phi_L), (h_I, \phi_I), (h_R, \phi_R)\}$ 
satisfying requirements (B1) and (B2).
It should be noted that the last comment in the first observation is taken into account for concluding (B2).
It follows from the definition of state variables that the length of curves of right states followed by $1$-shocks determines upper bounds of the bump height.
Moreover, the plateau regions in $2$-shock curves mentioned before imply that particles are homogeneously distributed in bumps of mixture fluids {\em for any height of tips}, as long as $a < h_R$ holds.
\par
We have left a comment about the relationship of speeds $s_1, s_2$ of shocks observed in Figure \ref{fig-locus_var_a} as the third observation.


\par
\bigskip
Here we compare our results with preceding works such as \cite{CBH, ZDBH}.
In all studies, results with $(h_L, \phi_L) = (1.0, 0.3)$ and $\rho_p = 2500$ are obtained.
According to \cite{ZDBH}, the bump structure by means of shocks in the above setting\footnote{
In \cite{ZDBH}, the coefficient $1/18a^2$ appeared in (\ref{wall}) is replaced by $1/2$, corresponding to $a = 1/3$. 
} is obtained as the triple of constant states $(U_L, U_I, U_R)$ with
\begin{equation*}
U_L \equiv (h_L, \phi_L) = (1.0, 0.3),\quad U_I \equiv (h_I, \phi_I) \approx (1.1619, 0.3620),\quad U_R \equiv (h_R, \phi_R) \approx (0.1, 0.3)
\end{equation*}
such that $(U_L, U_I)$ is followed by the $1$-shock with the speed $s_1\approx 0.4556$ and that $(U_I, U_R)$ is followed by the $2$-shock with the speed $s_2\approx 0.4956$.
On the other hand, according to our computations with $(a/h_0)^2 = 0.0001$, the bump structure (in the sense of Lax) can be constructed when $h_I \leq 1.1210$ and $\phi_I \leq 0.3903$ for $1$-shocks whose corresponding shock speed is $s_1 \approx 0.44927$.
In addition, according to our computations, the shock speed $s_2$ following the states $(h_I, \phi_I) = (1.1210, 0.3903)$ and $(h_R, \phi_R) \approx (0.1, 0.3)$ is $s_2 \approx 0.4930$, which approximately agree with results in preceding studies up to the small corrections of values.
It should be mentioned that the present model (\ref{system1}) is originally obtained in \cite{CBH}, which is slightly different from \cite{ZDBH}.
In \cite{CBH}, the entropy conditions for characterizing Lax's shock are incorrect, while our entropy conditions (\ref{1entropy})-(\ref{2entropy}) follow from general mathematical theory of shocks (cf. \cite{S}).
This gap induces small but nontrivial difference of the range of constant states admitting Lax's shocks.
\par
Finally we mention results in \cite{CBH} where shock structure with {\em given} right states $(h_R, \phi_R)$ is studied.
In particular, significantly small choice of $\phi_R$ is also mentioned, such as $0.002$ and $0.0005$.
Incorporating our present results with the preceding results, the shock structure in the sense of Lax is considered to be physically meaningful only if $a < h$ holds, which indicates that too small choice of $\phi_R$ can be meaningless for describing Lax's shocks when the effect of particle-wall interaction $W(h)$ is explicitly considered and the radius $a$ is inappropriately chosen, while it is not explicitly mentioned in \cite{CBH}.
It should be also mentioned that the choice of small $\phi_R$ does not deny the possibility of solutions of (\ref{system}) with the right state $(h_R, \phi_R)$ by means of other type of (elementary) waves, such as rarefactions or singular shocks which cannot be characterized by means of Hugoniot loci. 

\subsection{Effect of particle concentration on shock structure}
\begin{figure}[htbp]\em
	\begin{minipage}{0.33\hsize}
		\centering
		\includegraphics[width=6.0cm]{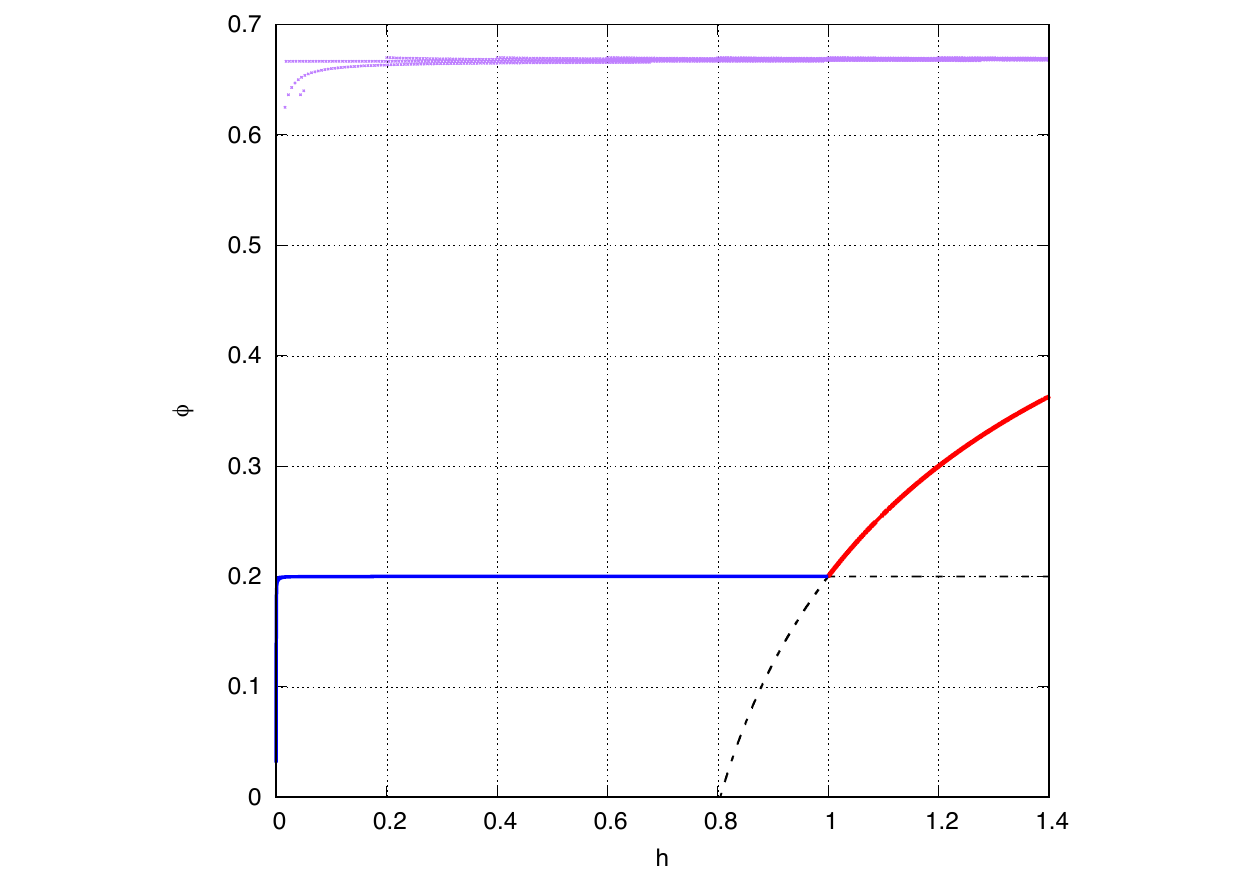}
	(a)
	\end{minipage}
	\begin{minipage}{0.33\hsize}
		\centering
		\includegraphics[width=6.0cm]{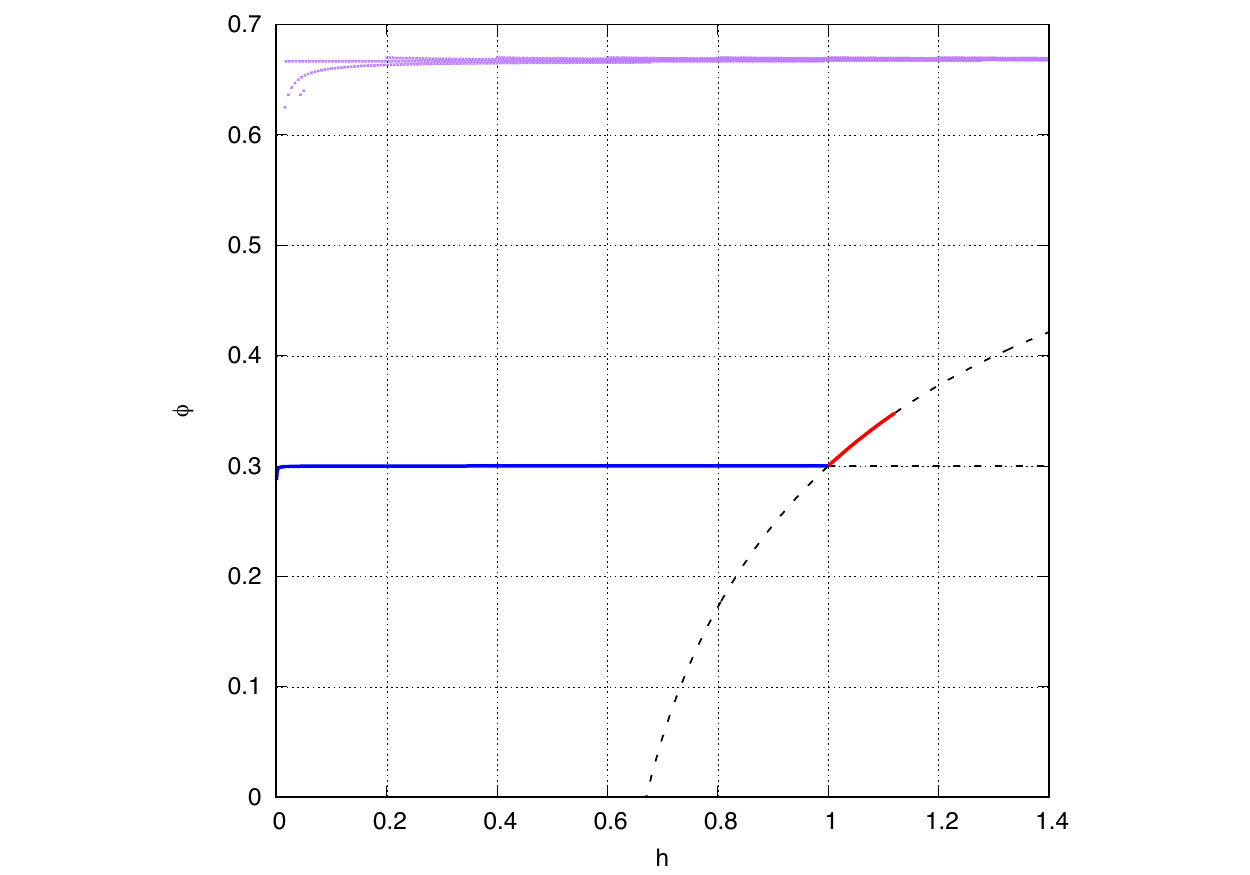}
	(b)
	\end{minipage}
	\begin{minipage}{0.33\hsize}
		\centering
		\includegraphics[width=6.0cm]{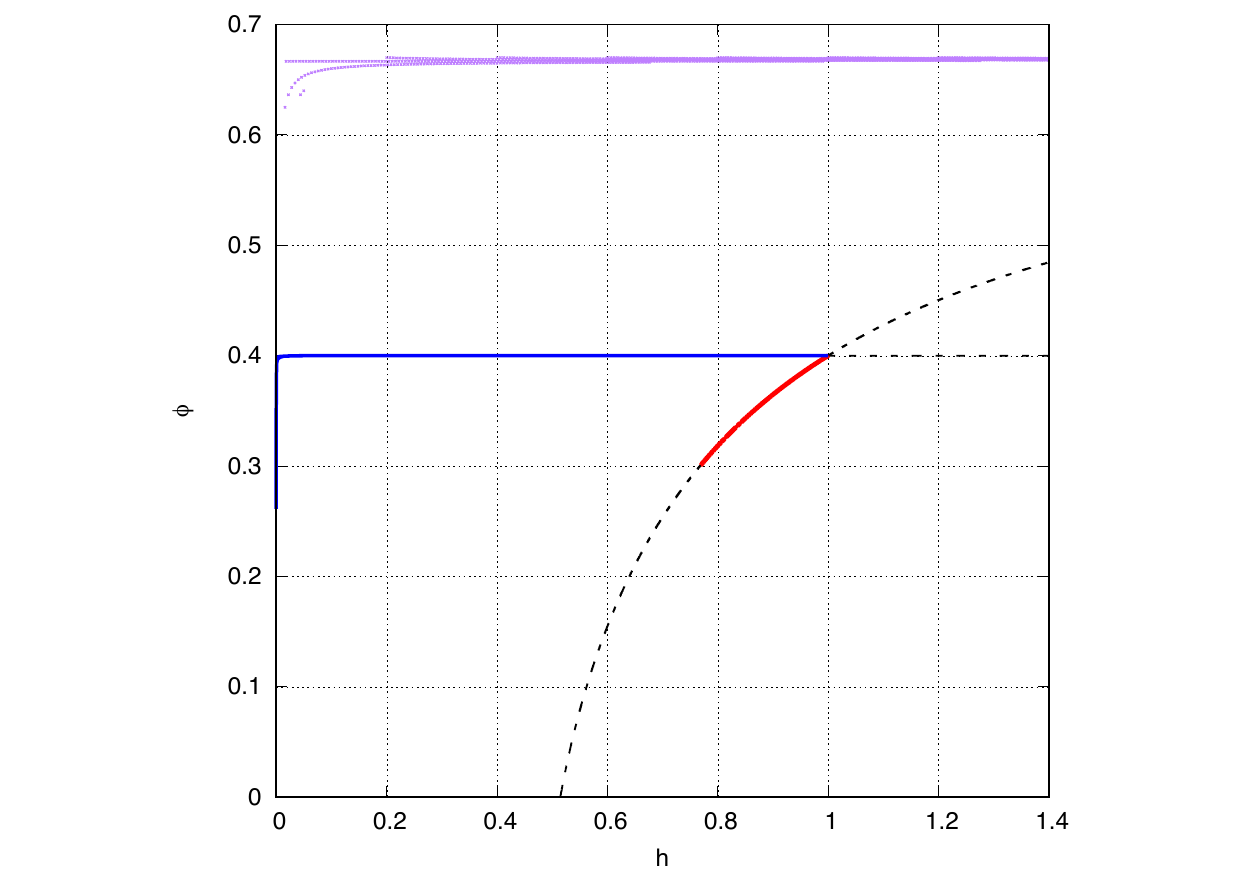}
	(c)
	\end{minipage}
	\caption{Hugoniot loci through $(h_L, \phi_L)$ with various $\phi_L$, $a=0.01$}
	\flushleft
	\label{fig-locus_var_phi}
	The basic description rule of figures is the same as previous figures.
	(a): $\phi_L = 0.2$. 
	(b): $\phi_L = 0.3$, which is the same as Figure \ref{fig-locus_sample}-(a).
	(c): $\phi_L = 0.4$. 
	In (c), the $1$-shock branch is distributed in $\{h < h_L\}$.
\end{figure}

Next, we study the effect of the initial concentration $\phi_L$ of particles on the shock structure.
At present we fix $a$ as $0.01$ and vary $\phi_L$, while all remaining parameters are not changed.
Figure \ref{fig-locus_var_phi} shows the Hugoniot loci with left states $(h_L, \phi_L) = (1.0, 0.2), (1.0, 0.3)$ and $(1.0, 0.4)$, respectively.
In any cases, two Hugoniot loci through $(h_L, \phi_L)$ are computed whose either side consists of right states connected to $(h_L, \phi_L)$ followed by $1$- or $2$-shocks.
An interesting observation here is that there is a collection of right states $(h,\phi)$ connected from $(h_L, \phi_L)$ followed by a $1$-shock satisfying $h_L < h$ and $\phi_L < \phi$ {\em  when $(h_L, \phi_L) = (1.0, 0.2)$ and $(1.0, 0.3)$, while it is not observed when $(h_L, \phi_L) = (1.0, 0.4)$}.

\par
Robustness of Lax's shock structure ensures that the above observation persists under small perturbations of $(h_L, \phi_L)$ with an appropriate choice of $(h_I, \phi_I)$, which can be confirmed at least in the numerical sense\footnote{
In the present system, the regularity of the Jacobian matrix of $G$ in (\ref{zero-conservation}) with respect to $(s,n)$, continuity of eigenvalues $\lambda_1, \lambda_2$, Lax's entropy condition and the genuine nonlinearity of characteristic fields as well as the Implicit Function Theorem for $G({\bf x}, {\bf p})=0$ ensure the robustness.
}.
Our next interest is then {\em the limit of $(h_L, \phi_L)$ where the bump structure generated by Lax's shocks persists}.
In the present study we fix $h_L = 1$ throughout and change $\phi_L$. 
Again $a$ is fixed as $0.01$.
Numerical investigations based on the bisection method yield the following observation, which indicates the limitation of generating bump structure.
\begin{description}
\item[(Limit)] When we fix $h_L = 1.0$ and $a=0.01$, there is an intermediate state $(h_I, \phi_I)$ with $h_I \leq 1.4$ followed by a $1$-shock if and only if $\phi_L \leq \phi_{L,\ast}$, where $\phi_{L,\ast} \approx 0.3314575$.
\end{description}

\subsection{$2$-parameter dependence on shock structure}
The observation (Limit) implies that the initial concentration $\phi_L$ of particles therefore affect the generation of bumps, in particular the limitation of bump structure generation.
On the other hand, observations in Figure \ref{fig-locus_var_a} indicate that the generation of bump structure by means of Lax's shocks depends not only on $\phi_L$ but also $a$.
It is then worth investigating the distribution of parameters $(\phi_L, a)$ where the bump structure can be generated.
According to Figure \ref{fig-locus_sample}, collections of right states $(h_R, \phi_R)$ from $(h_I, \phi_I)$ followed by $2$-shocks satisfying (B2) are typically generated, and hence investigation of the existence of intermediate states satisfying (B1) provides the necessary, and possibly sufficient environment for generating the bump structure. 
To this end, we fix $h_L = 1.0$ as before and vary both $\phi_L$ and $a$. 
Hugoniot loci including states satisfying (B1) with various $\phi_L$ and $a$ are mainly considered next.

\begin{figure}[htbp]\em
	\begin{minipage}{0.33\hsize}
		\centering
		\includegraphics[width=6.0cm]{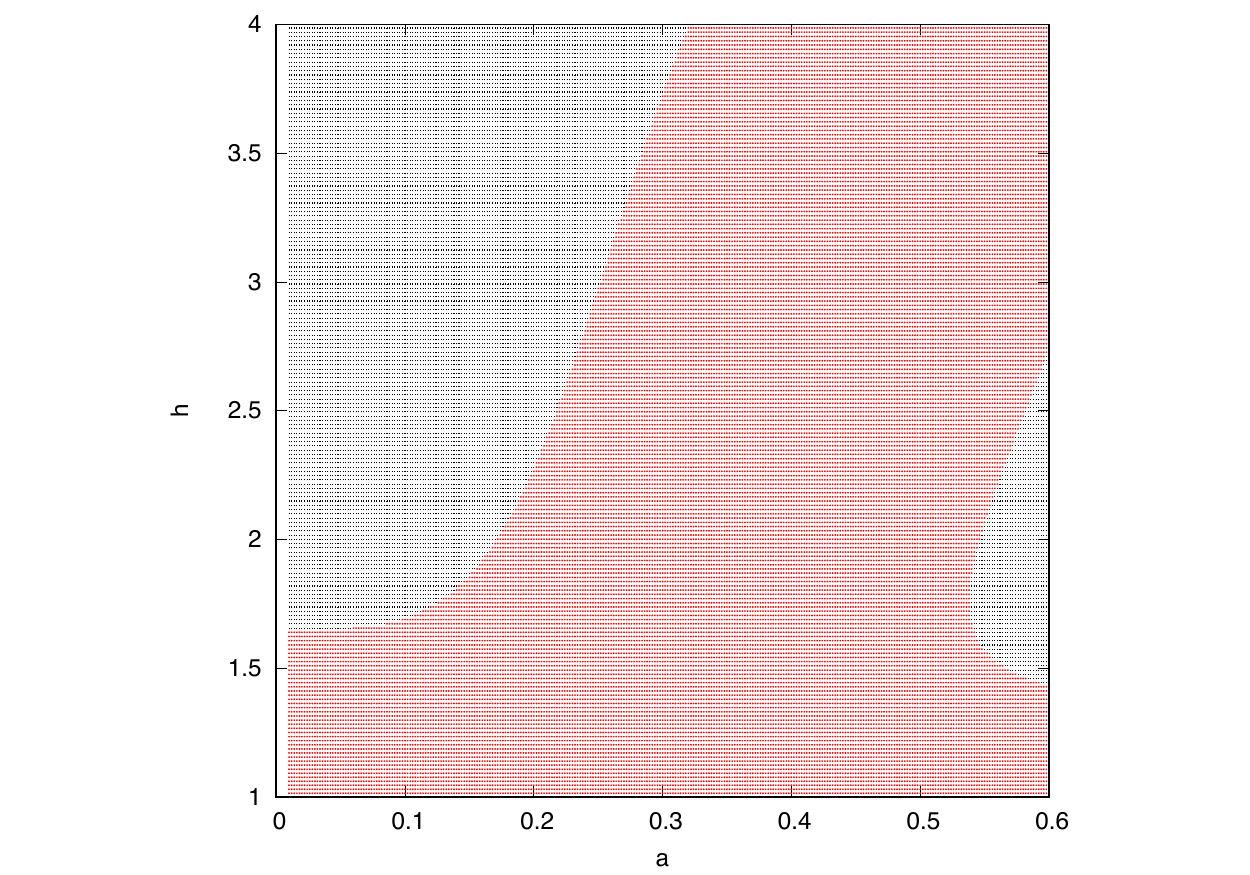}
	(a)
	\end{minipage}
	\begin{minipage}{0.33\hsize}
		\centering
		\includegraphics[width=6.0cm]{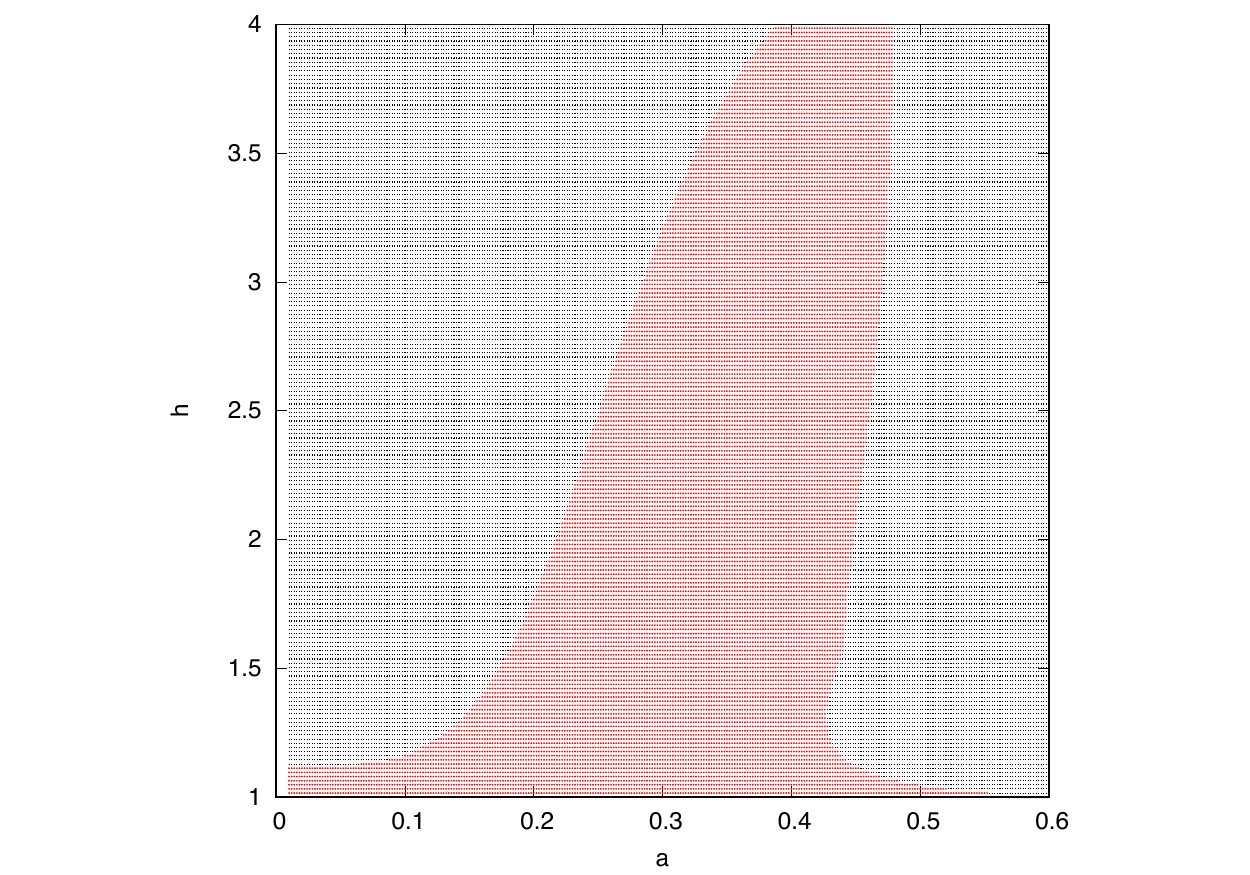}
	(b)
	\end{minipage}
	\begin{minipage}{0.33\hsize}
		\centering
		\includegraphics[width=6.0cm]{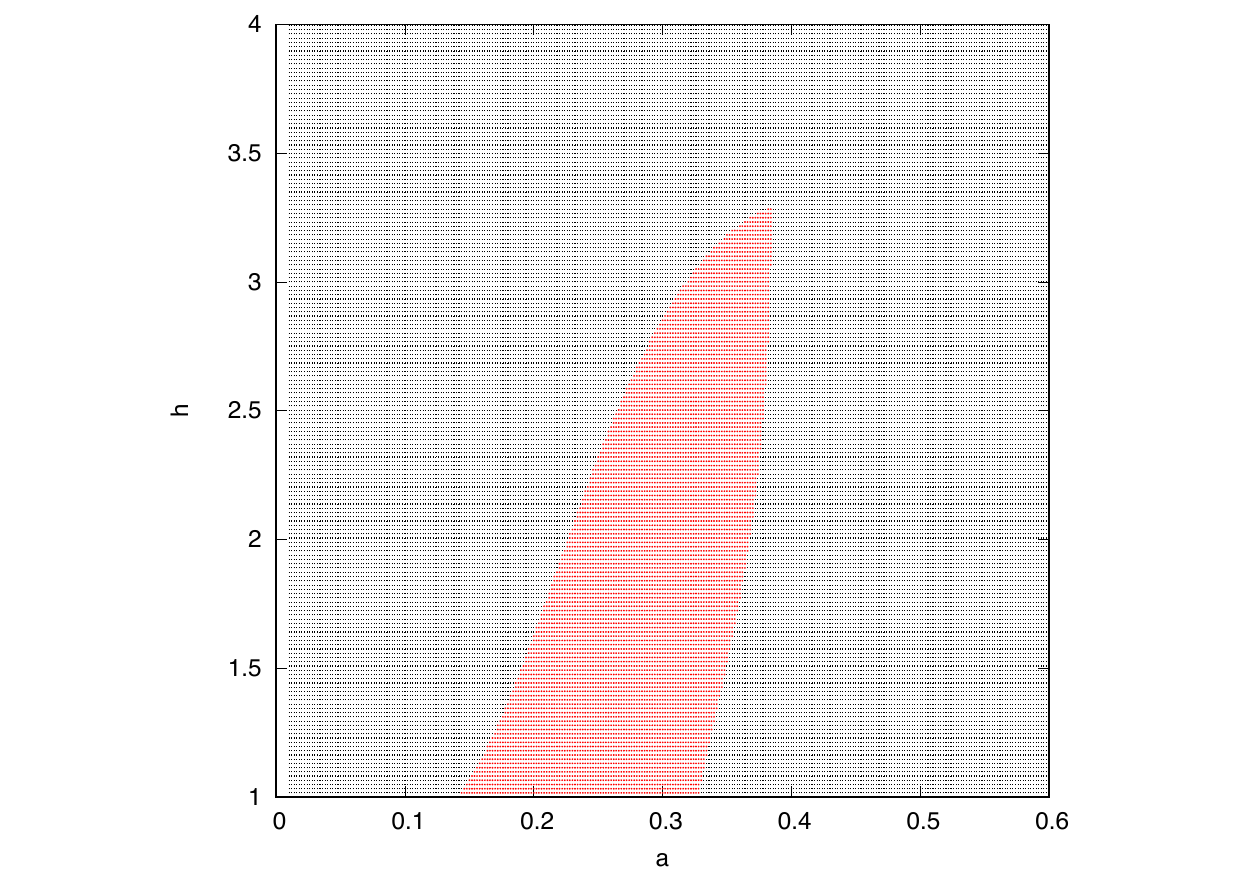}
	(c)
	\end{minipage}
	\caption{The region of $(a, h)$ admitting $1$-shocks satisfying (B1) for fixed $\phi_L$}
	\flushleft
	\label{fig-1shock_a_h}
	In all figures, the projection of the Hugoniot locus on $(a,h)$-plane through $(h_L, \phi_L)$ possibly including right states followed by $1$-shocks are drawn.
	As in previous figures, the red region denotes the collection of right states followed by $1$-shocks, while the black region denotes the collection of right states violating entropy conditions.
	The horizontal axis represents $a\in [0.01, 0.6]$ and the vertical axis represents $h\in [1,4]$.
	(a): $\phi_L = 0.2$. 
	(b): $\phi_L = 0.3$. 
	(c): $\phi_L = 0.4$. 
\end{figure}

First, distribution of $(a, h)$ with various $\phi_L$ such that $1$-shocks are admitted is summarized in Figure \ref{fig-1shock_a_h}, where the range of $h$ is restricted to $h>1$.
Looking at results for fixed $\phi_L$, we first observe that the upper bound $h_{\max}$ of $h > h_L$ admitting $1$-shocks converges to a certain value as $a \to 0$ when $\phi_L$ is relatively small.
For example, when $\phi_L = 0.3$, the upper bound is $h_{\max} \approx 1.121185$, while $h_{\max}\approx 1.6519053$ when $\phi_L = 0.2$ and $h_{\max}\approx 0$ when $\phi_L = 0.4$.
This observation indicates that the bump generation mechanism is robust among different particles with sufficiently small radii.
For relatively small $\phi_L$, say $0.2$ (Figure \ref{fig-1shock_a_h}-(a)), $1$-shocks are admitted for a large range of $a$, while the range of $h$ admitting $1$-shocks is significantly changed as $a$ increases.
When $a$ becomes larger, the range of $h$ admitting $1$-shocks also becomes large, which is significant in the middle particle size range, say $a\in [0.3, 0.5]$.
When $a$ becomes much larger, say $a \geq 0.54$, the region \lq\lq bifurcates" into two subregions.
In particular, only small and considerably large bumps can be generated, but the region of \lq\lq large" $h$ is admitted in a thin range of $a$. 
The qualitatively similar feature can be also seen in environments with a little larger $\phi_L$, say $0.3$ (Figure \ref{fig-1shock_a_h}-(b)).
On the other hand, when $\phi_L = 0.3$, we also see that only low height bumps can be generated whose height converges to $0$ as $a$ becomes larger and larger.
Furthermore, for much larger $\phi_L$, say $0.4$  (Figure \ref{fig-1shock_a_h}-(c)), bump structure is generated only for middle particle size range, say $a\in [0.2, 0.4]$.
This result is compatible with the observation (Limit).
\par

\begin{figure}[htbp]\em
	\begin{minipage}{0.5\hsize}
		\centering
		\includegraphics[width=6.0cm]{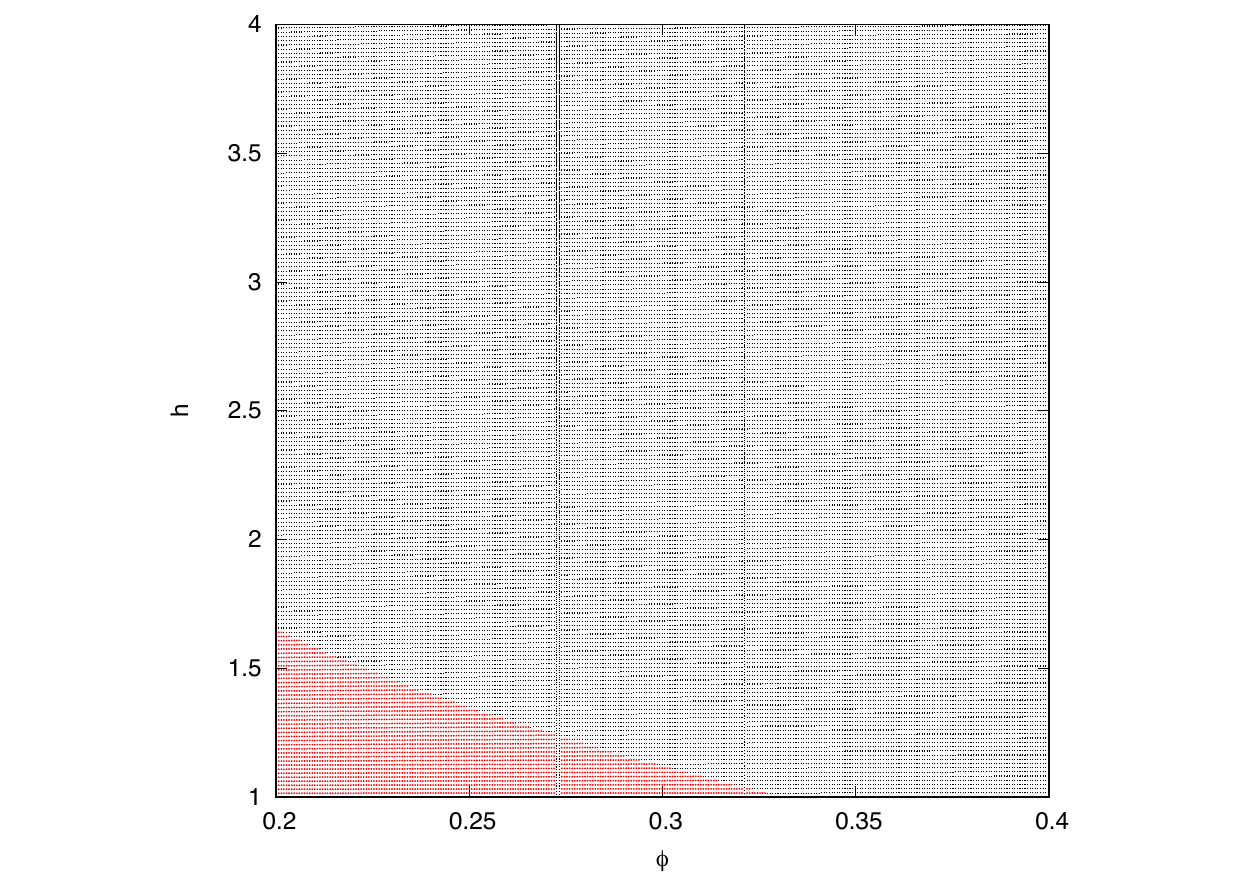}
	(a)
	\end{minipage}
	\begin{minipage}{0.5\hsize}
		\centering
		\includegraphics[width=6.0cm]{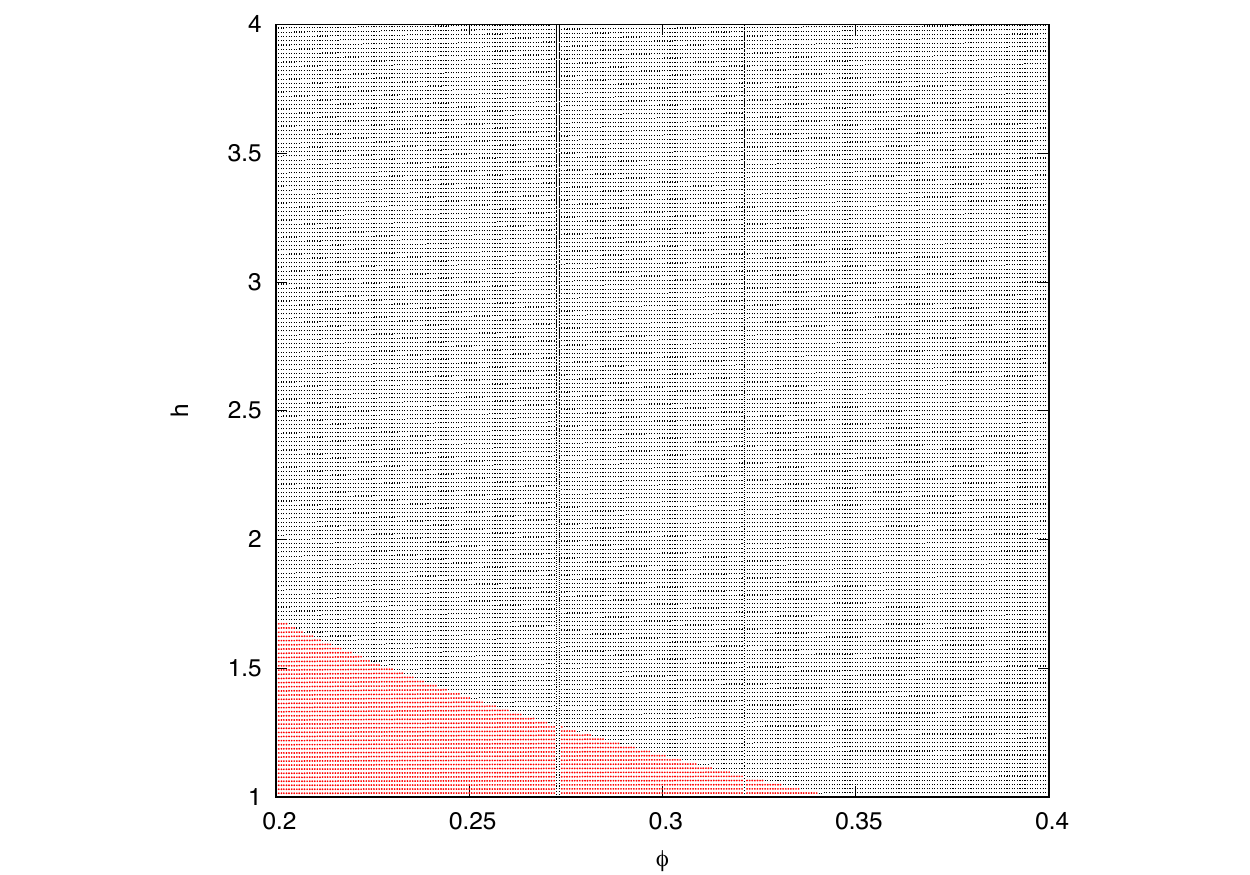}
	(b)
	\end{minipage}\\
	\begin{minipage}{0.5\hsize}
		\centering
		\includegraphics[width=6.0cm]{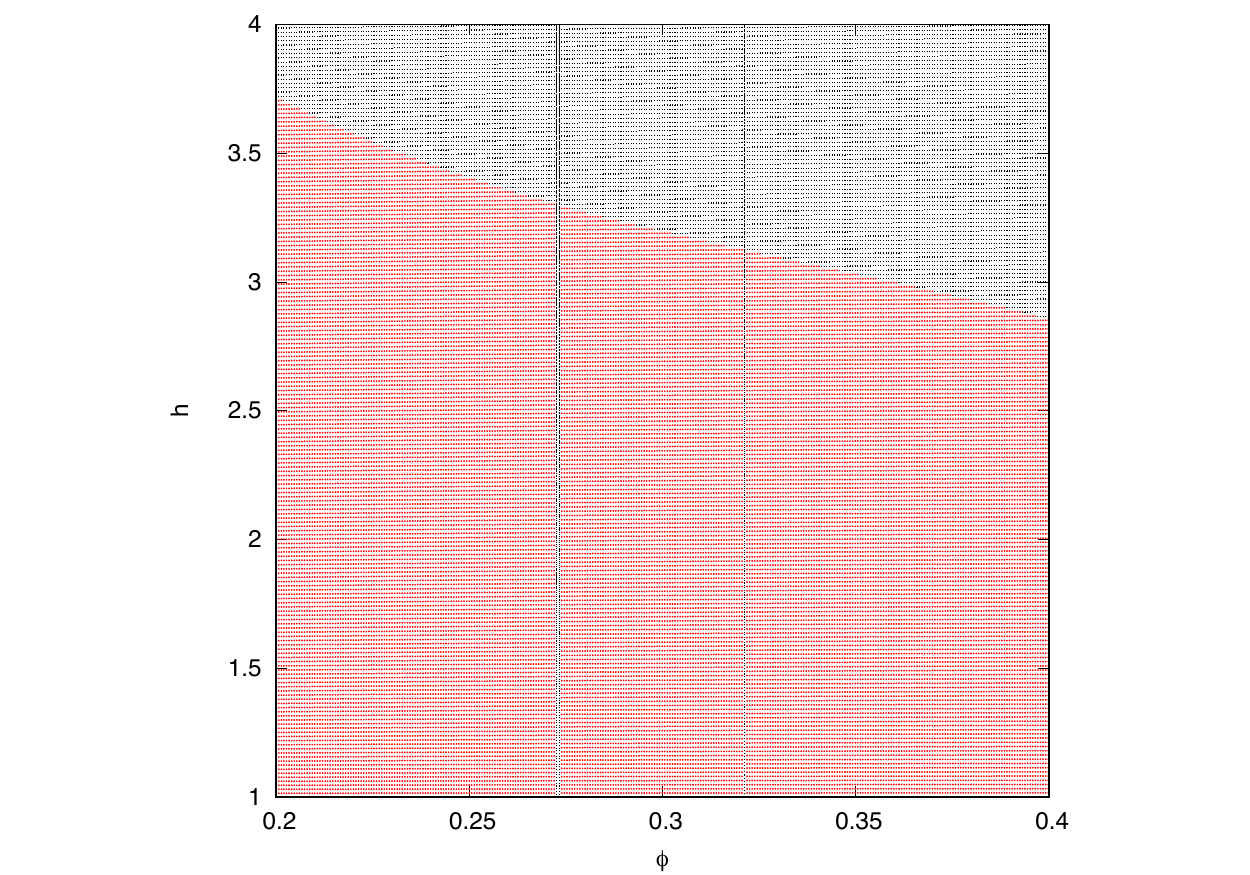}
	(c)
	\end{minipage}
	\begin{minipage}{0.5\hsize}
		\centering
		\includegraphics[width=6.0cm]{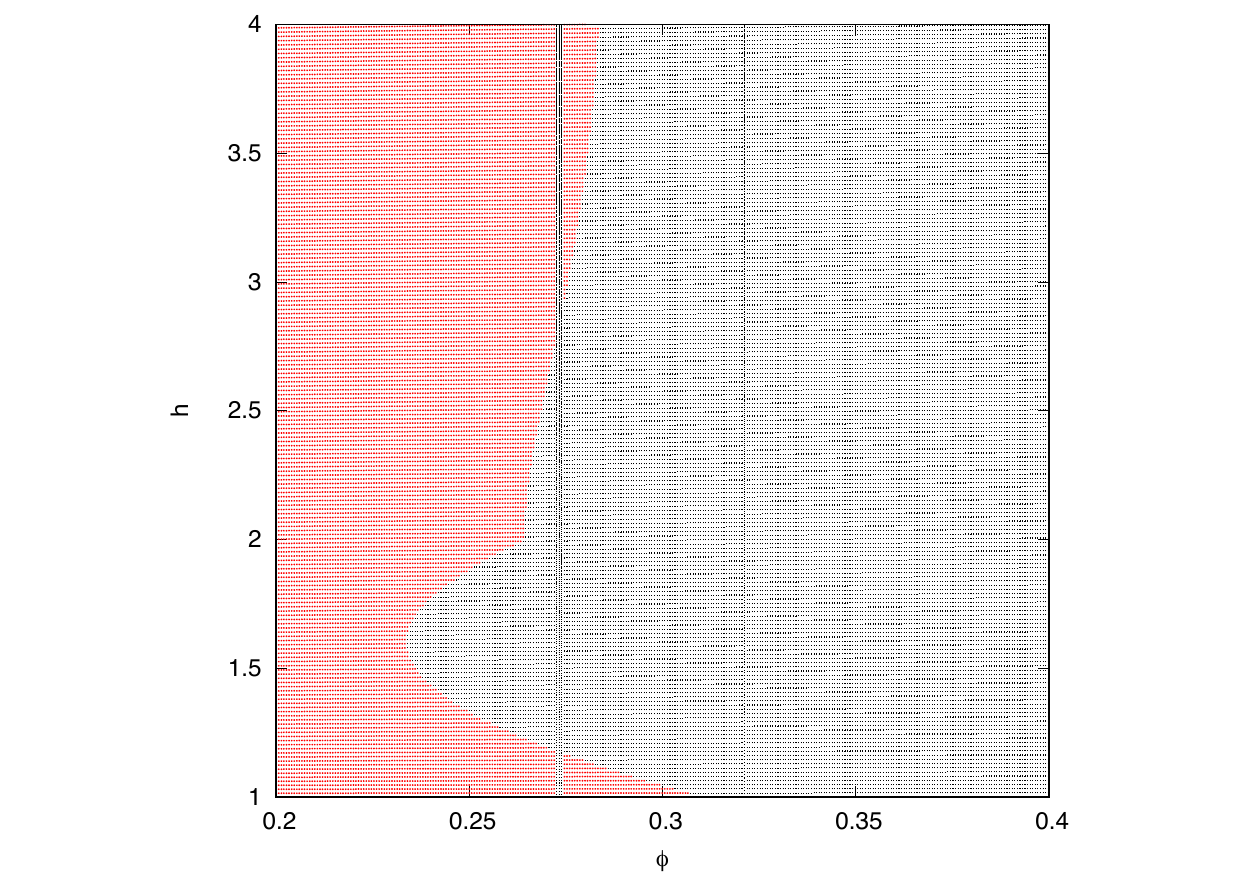}
	(d)
	\end{minipage}
	\caption{The region of $(\phi_L, h)$ admitting $1$-shocks satisfying (B1) for fixed $a$}
	\flushleft
	\label{fig-1shock_phi_h}
	The basic description rule of figures is the same as Figure \ref{fig-1shock_a_h}.
	while the present figures represent the projection of the Hugoniot loci on $(\phi,h)$-plane.
	The horizontal axis is replaced by $a\in [0.2, 0.4]$.
	(a): $a = 0.01$. 
	(b): $a = 0.1$. 
	(c): $a = 0.3$. 
	(d): $a = 0.5$. 
\end{figure}

Comparing all figures in Figure \ref{fig-1shock_a_h}, the region admitting $1$-shocks looks smaller and smaller as $\phi_L$ increases in a monotonous manner, while it is indeed nontrivial since contributions of $a$ and $\phi_L$ in the problem (\ref{zero-finding}) is nonlinear.
We then study the distribution of $(\phi_L, h)$ with various $a$ such that $1$-shocks are admitted as the next step, which is summarized in Figure \ref{fig-1shock_phi_h}, where the range of $h$ is restricted to $h>1$ like Figure \ref{fig-1shock_a_h}.
As seen in Figure \ref{fig-1shock_phi_h}, the region where $1$-shocks are admitted monotonously shrinks as $\phi_L$ becomes large, for small $a$ such as $0.01$ and $0.1$.
When $a = 0.01$, the upper bound of $\phi_L$ where $1$-shocks are admitted is $\phi_{L,\ast}$ as computed before.
The monotonously shrinking behavior can be also seen in the middle size particle, say $a=0.3$, but the region of $h$ admitting $1$-shocks itself is significantly large for all $\phi_L$ under consideration.
Compared with Figure \ref{fig-1shock_a_h}, this feature reflects the significant increase of the region for $a$ around $0.3$.
When $a$ is considerably large, say $a=0.5$, the region of the presence of $1$-shocks depends on $\phi_L$ in a nontrivial manner.
As in the previous observation, the region \lq\lq bifurcates" at $\phi_L \approx 0.23397$ and the boundary of the region in $(\phi_L, h)$-plane becomes non-smooth at $\phi_L \approx 0.2642$.
As $\phi_L$ becomes much larger, say $\phi_L \geq 0.32$ with $a=0.5$, no $1$-shocks of our interests can be generated, as can be seen in Figure \ref{fig-1shock_a_h}.

\section{Concluding remark}
\label{sec:5}
%
%

In this paper, we have studied the bump structure of 
particle laden flows through the simplified model of conservation laws.
The dependence of radius and initial concentration of particles in films on ridge is mainly focused taking the  particle-particle interaction and particle-wall interaction into account, although the latter interaction is intrinsically considered only in an asymptotic sense in preceding studies.
The bump structure is considered by means of composite waves consisting of Lax's shocks.
In the present study, we have observed the nonlinear dependence of radius and concentration of particles on bump structure of shocks.
For relatively small radii of particles so that the treatment of particle laden flows as continua is valid, the bump structure is generically admitted with a non-trivial but limited range of heights which depends on the initial concentration of particles. 
As the radius becomes large, the bump formation mechanism becomes complicating and significantly high bump can be also generated, while bumps with moderate height are prevented.
For fixed radii, the range of bump heights is monotonously reduced as the initial concentration of particles increases when the radius is relatively small.
When radius becomes large, the mechanism of bump generation behaves in a nonlinear manner as a function of initial concentration.
Unlike the preceding works such as \cite{CBH, ZDBH}, the present study extracts the genuine contribution of particle characteristics to morphology of particle laden flows, possibly with their physical reliability in terms of mathematical arguments based on the theory of conservation laws and objects such as Hugoniot loci.


%
%


\par
\par
In Murisic et. al. \cite{MPPB}, a modified model including the effect of shear-induced migration is proposed, where the importance of the particle size considerations we have also mentioned is emphasized in their experiments, especially in the transient (well-mixed) regime.
Several numerical simulations by e.g. \cite{WM} show the existence of the bump (double-shock) and singular shock structure with relatively high initial particle concentrations $\phi_L = 0.5$.
The reduced model for characterizing their targeting objects by means of shocks is different from our model considered here, and comparison of solution structures with those in the present model are next directions of our work for extracting similarity and difference of results towards the study of the intrinsic contribution of, say shear-induced migration.

\section*{Acknowledgements}
This work was supported by JSPS Grant-in-Aid for Scientific Research (C) (No. JP18K03437). 
KM was partially supported by Program for Promoting the reform of national universities (Kyushu University), 
Ministry of Education, Culture, Sports, Science and Technology (MEXT), Japan, World Premier International Research Center Initiative (WPI), MEXT, Japan and JSPS Grant-in-Aid for Young Scientists (B) (No. JP17K14235). 


\end{document}